\documentclass[]{agujournal2019mod}
\newcommand{\ltsim}{\raisebox{-1.0ex}{$\stackrel{\textstyle<}{\sim}$}}
\def\kms{km~s$^{-1}$}
\def\al{Alfv\'{e}n}

\def\yohkoh{{\sl Yohkoh}}

\def\hinode{{\sl Hinode}}

\def\p78{{\sl P78-1}}

\def\soho{{\sl SOHO}}
\def\trace{{\sl TRACE}}
\def\stereo{{\sl STEREO}}
\def\sdo{{\sl SDO}}
\def\iris{{\sl IRIS}}

\def\caii{Ca~{\sc ii}}

\def\al{Alfv\'{e}n}

\def\kms{km~s$^{-1}$}

\def\etal{et~al.}

\begin{document}
%

\title{Fine-Scale Features of the Sun's Atmosphere: Spicules and Jets}

\correspondingauthor{Alphonse C. Sterling}{alphonse.sterling@nasa.gov}

\authors{Alphonse C.~Sterling}
\affiliation{}{Marshall Space Flight Center, Huntsville, AL 35812, USA}

\begin{abstract}
We present an overview of fine-scale features in the Sun's atmosphere, with a focus on spicules and jets.
We consider older and newer observations and theories for chromospheric spicules and coronal jets.  We also
consider the connection between these features and some other solar atmospheric phenomena.  We then discuss
the possibility that there is a continuum of jet-like features ranging from spicules to large-scale
CME-producing eruptions, all driven by similar magnetic processes operating on differing corresponding size
scales.  Future observational and theoretical studies will help clarify further the nature of these solar
events, and elucidate possible connections between them.
\end{abstract}


\section{Introduction}
\label{sec-introduction}

A striking aspect of a typical image of sufficient resolution showing the Sun's atmosphere above the photosphere 
is the amount of structure present.  Frequently this structure is in the form of transient and dynamic features that
resemble geysers, in that they grow to be longer than (sometimes many times longer than) their widths.  These are 
jets of solar plasmas that shoot out from near the photosphere into higher atmospheric layers.  They are seen
on the size scale of less than an arcsecond to tens of arcseconds.  These fascinating jet-like features, which 
make up key components of different regions of the solar atmosphere, are the topic of this Section.

We will anchor our discourse on two primary jet features: Spicules and coronal jets.  Spicules are part of the
chromosphere.  They are comparatively small features, with 
widths near the limits of most ground-based instruments and extending only into the bottom of the corona 
(widths $\sim$0$''.5$, heights $\sim$5$''$---$10''$).  They have been 
primarily observed at visible wavelengths such as H$\alpha$ and \caii.    Chromospheric spicules
have lifetimes of a few minutes, and occur in vast numbers over essentially all of the Sun's surface and seemingly in all solar 
regions (perhaps avoiding sunspots). Thus they are a key constituent of the solar chromosphere. Similar 
features appear at UV and EUV wavelengths.  For a long time it was
not known whether these made up a hot component of (chromospheric) spicules, or if they are entirely different 
entities; thus these features usually are described with an adjective, that is, UV spicules or EUV spicules, to 
differentiate them from the chromospheric
spicules..  Recent investigations however now show that at least some of the UV spicules are hotter components of chromospheric
spicules (\S\ref{subsec-recent spicules}). Figure~1 shows examples of spicules at the solar limb observed in
\caii\ by the \hinode\ spacecraft.

\begin{figure}[ht!]
\hspace*{-2.0cm}\includegraphics[angle=0,scale=1.00]{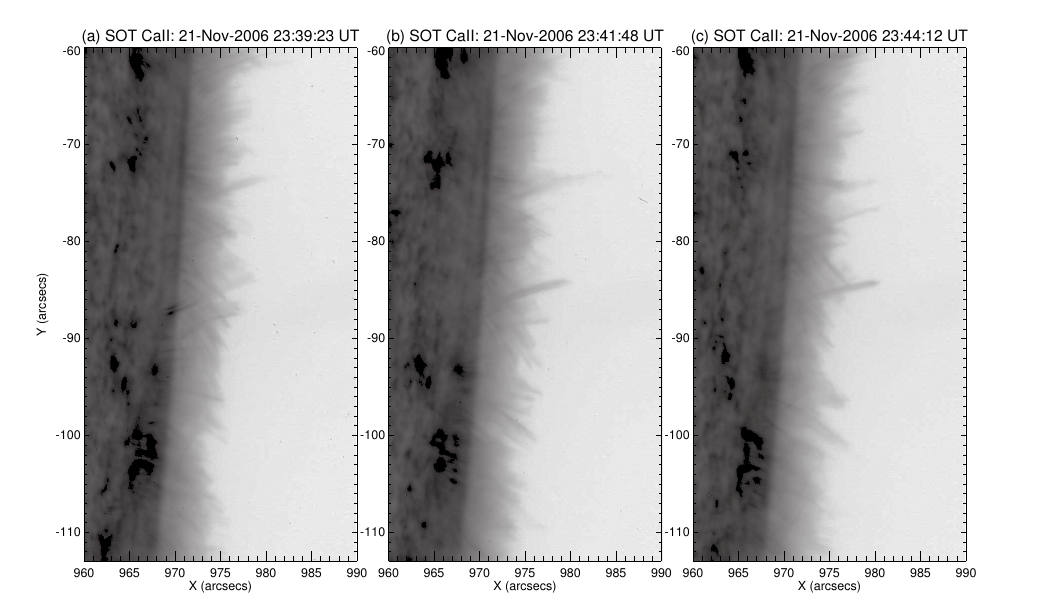}\vspace{0.0cm}
\caption{Examples of chromospheric spicules observed at the solar west limb with the \hinode\ spacecraft in \caii.  Dynamic
changes can be seen in many of the spicules that reach across the limb over the course of panels (a)---(c).  Intensity
colors are reversed, so that bright features appear dark and vice versa.  Data from
this period were analyzed by \citeA{depontieu.et07a} and by \citeA{zhang.et12}. \label{fig1}}

\end{figure}

Coronal jets, as the name indicates, are coronal features, primarily seen at EUV and X-ray wavelengths; thus they 
are largely only observable from space (except for their chromospheric components).  They have 
size scale larger than those of spicules, with widths $\sim$5$''$---$10''$, and heights of a few 10s or 100's of 
arcseconds.  They
typically last for ten minutes or so, and occur in coronal holes, quiet Sun, and in the periphery 
of active regions (ARs).  Figure~2 shows examples observed in X-rays in a polar coronal hole (Fig.~2(a)),
and near an AR (Fig.~2(b)).

Following a discussion of these two basic features (spicules and jets), we will briefly discuss: features of 
size scales in-between spicules and jets, the question of whether there is a connection between features in 
these two broad categories, and the relationship of jet-like features with other solar activity.

\begin{figure}[ht!]
\hspace*{-1.0cm}\includegraphics[angle=0,scale=0.85]{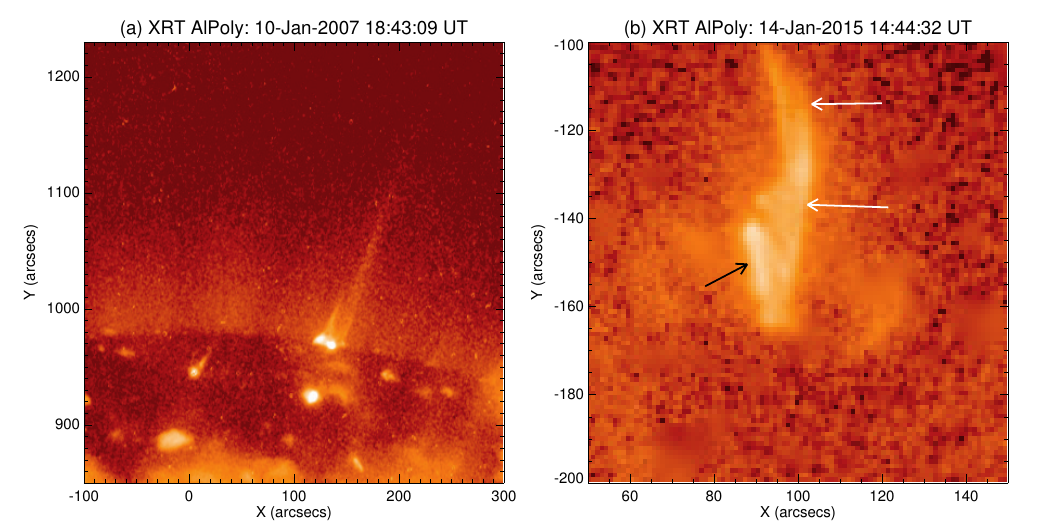}\vspace{0.0cm}
\caption{Examples of X-ray images of solar coronal jets observed with the \hinode/ X-ray telescope (XRT), with (a) showing the north
polar coronal hole limb region (taken with the XRT ``Al Poly" filter), and (b) showing a closeup of a jet in an active region on the 
disk (taken with the ``Be thin" filter); white arrows point to the spire.  The typical morphology for the jets is a spire shooting 
away from the Sun, with a bright base at the bottom of the spire.  Often one side of the base is particularly bright (e.g., the 
left side of the base in (b); black arrow), a feature we refer to as a ``jet-base bright point" (following \citeA{sterling.et15}). 
The jets in (a) were studied by \citeA{cirtain.et07}, and that in (b) was studied in \citeA{sterling.et17}. \label{fig2}}
\end{figure}

\section{Solar Spicules}
\label{sec-spicules}

Because they are observable in the visible portion of the spectrum, spicules have been studied for a long time.
They were named by \citeA{secchi877}, and studied since then by many workers.  They are always on the Sun,
unlike sunspots and other active-region features visible from the ground.  Their near-resolution-limit
visibility left even their basic properties a challenge to define, likely adding to their allure in some circles.   
Study of spicules however seems to have hit a trough during the decades of the 1970s thorough 
the early 2000s, over which time space-based observations were popularizing other areas of solar research,
areas that had previously been totally unaccessible from the ground, such as observations at UV and shorter 
wavelengths.  Although
some spicule-like features were studied in the UV during this period (e.g., \citeA{dere.et83}), many of the
space-based observations were of spatial resolution too coarse to resolve spicules.  Improvements
in ground-based capabilities, direct observations of the chromosphere from space with the \hinode\ 
satellite, and the advent of continuous, high-resolution observations of the lower corona and the transition
region, resulted in a ``spicule-observation renaissance'' from about 2000.  We consider spicules observations 
during the earlier period, and then the later renaissance period.

The literature on spicules is extensive, and we are not attempting a full review here. \citeA{beckers68} and \citeA{beckers72} are two 
highly cited review papers covering the earlier studies.  \citeA{sterling00} updates some of the observational discussion
and reviews theoretical ideas for spicules since the Beckers papers, with an emphasis on results of numerical 
modeling.  \citeA{tsiropoula.et12} provide an updated review emphasizing observations, and \citeA{zaqarashvili.et09} review
waves in and on spicules.

\subsection{Earlier Spicule Observations}
\label{subsec-early spicules}

Spicules were detected with the advent of H$\alpha$ observations of the chromosphere in the late 1800s and early 1900s,
and were also studied in white light eclipse photographs.
Given adequate atmospheric seeing and sufficient spatial resolution, they appear prominently in emission 
against the dark background of space at the solar limb.  Their presence gives a fine-scale 
non-uniformity to the H$\alpha$ limb.  When those early observations were of sufficient cadence, individual spicules 
were seen to move out from the photosphere and reach into the corona, and then to either fall back to the photosphere
or to fade before falling.

Due to their small size that rendered them near the edge of detectability with instruments of those times, it was a challenge
even to characterize consistently the properties of spicules.  Those and other works 
(e.g., \citeA{rush.et54}, \citeA{lippencott57}) resulted in some average 
values for spicules, such as maximum heights of 6500---9500~km, widths of $\ltsim$300---1500~km 
(e.g., \citeA{nishikawa88}, \citeA{lynch.et73}), and lifetimes of 1---10~min \cite{lippencott57,bray.et74}.  Other studies 
yielded spicule temperatures of 5000---15{,}000~K and
densities of $\sim$3$\times 10^{-13}$~g~cm$^{-3}$; there was much discussion of spicule rise velocities,
with a typical quoted value being $\sim$25~\kms (see summaries in, e.g., \citeA{beckers68,beckers72}).  

There has also consideration of whether these spicules
were observed to fall back to the solar surface.  The upshot is that while both rising and falling motions were reported, some 
percentage fade from view (in the chromospheric line of the observation) after reaching their maximum heights.  That 
percentage varies a large amount, perhaps
depending on the region in which the spicules occurred and/or the specific observation circumstances.  For example, 
\citeA{lippencott57} report that (only) 58\% of her sample of 77 spicules appear to descend or fade from the top, 
while \citeA{pasachoff.et09} report that $\sim$70\% of 40 QS spicules they observed fade.
\citeA{suematsu.et95} report that 80\% could be followed through complete up-and-down cycles for spicules 
they observed on-disk in a region of enhanced network.  Basic 
considerations regarding mass flux however show that something like 99\% of the material that goes upward in spicules must return to 
the surface.  That is, at least most of the fading spicule must return to the surface. This is because a calculation using the 
above properties for spicules, and assuming that, say, 30\% of the upward material escapes the 
Sun, results in an upward mass flux $\sim$100 times that observed in the solar wind.  This led to the suspicion that
many spicules might undergo sufficient heating during their assent to stop emitting in visible chromospheric lines.
And comparing the upward mass flux with observed UV downflows in the magnetic network helped fuel this
suspicion \cite{pneuman.et78}.

To understand the driving source of spicules it was important to know what was going on at their bases as they
started to rise upward.  That however was not (and still is not even today) trivial to address. A key difficulty
is that spicules are best seen at the limb.  But since they only live for a few minutes, it
is not possible to follow the same spicule seen at the limb as it rotates onto the solar disk.  (So, for example, if 
we observe a spicule at the east limb, it will be gone in five minutes, and so when the base of that particular spicule is on
the disk --- several hours or a day later --- the spicule, and the fine-scale magnetic flux arrangement at its base, will 
be long gone.)  Moreover, in contrast
to spicules always emitting when viewed at the limb, disk observations show spicule-like features (i.e., of similar
physical dimensions and roughly similar lifetimes) that were either absorbing or emitting: these features are
respectively called dark and bright mottles, with use of the new term (``mottle'') emphasizing that a one-to-one correspondence
with spicules had not been established unambiguously.  (NB. Some workers use the term spicules instead of mottles, e.g.\
\citeA{zirin88}.) Near-limb mottles were seen as limb spicules in careful studies \cite{koutchmy.et71}. 
Nonetheless, a key finding from H$\alpha$ on-disk studies of the spicule-like mottles is that they are rooted in
the magnetic network.  This supports that the spicules have a strong magnetic connection, in that they are at the
least channeled along the field.  Features resembling mottles that are more horizontal than vertical are referred to as
{\it fibrils} \cite{foukal71a,foukal71b,zirin88}.

Meanwhile, observations in UV \cite{dere.et83} showed features 
that resemble spicules (or macrospicules), in that they grow in length to become long and thin like spicules, 
but are much hotter ($\sim$10$^5$~K) than the chromospheric spicules.  And an unanswered question was that of 
the relationship between these {\it UV spicules} and normal (i.e.\ chromospheric) spicules.

Also, some EUV features \cite{bohlin.et75,withbroe.et76,dere.et89,harrison.et97,xia.et05,loboda.et19}  were described as 
larger-than-normal spicules, and hence called {\it macrospicules}.  \citeA{moore.et77} and \citeA{labonte79}
identified some macrospicules with miniature filament eruptions (for a different view, see \citeA{wang.et00}).
Several observations support that macrospicules spin \cite{pike.et98,sterling.et10a,kamio.et10}.

\subsection{Earlier Spicule Models}
\label{subsec-early spicule models}

Because the roots of on-disk spicule-like features were seen to congregate at the boundaries of the
magnetic network, it was clear very early on that spicules are dependent upon the magnetic field.
But what drives them?  Whatever it is, it must be a very common aspect of the lower solar atmosphere
in order to produce spicules in the observed copious quantities.  

Also, the mechanism must be able to supply sufficient amounts of energy to power the spicules.  A simple 
calculation shows that the required amount of gravitational energy, based on the average numbers presented above, is
$\sim$10$^{24}$---$10^{25}$~erg for an individual spicule (e.g., \citeA{sterling00}).  This energy can easily be supplied 
by photospheric motions, given that individual photospheric granules have kinetic energy flux about two-orders 
of magnitude larger than this.  On the other hand, the rate at which this energy must be supplied
into the area of the base of the spicule (assuming a lifetime of 5 minutes and a radius of 300 km, say) is 
$10^6$---$10^7$~erg~cm$^{-2}$~s$^{-1}$; this is a substantial value, in the sense that it is comparable
to the energy fluxes required to heat the corona.  If however we assume that spicules cover only about
1\% of the Sun's surface (which seems reasonable, based on the density of mottles on the disk), then 
these energy fluxes would {\it not} be sufficient for coronal heating.

This reasoning has been used in the past to argue that spicules are not capable of heating the corona.
More recent studies however (e.g., \citeA{falconer.et03,zaqarashvili.et09,moore.et11}) hold that the 
energy from spicules might indeed
be large enough to power the corona if one takes into consideration the energy in spicule oscillations, 
e.g.\ \al\ waves, which some recent observations suggest is substantial.  And in fact the question of whether spicules
heat the corona is a topic still under active debate \cite{depontieu.et11,madjarska.et11,klimchuk12,klimchuk.et14,bradshaw.et15}.

There are several excellent summaries of discussions of spicule models up to the early 1970s (e.g., 
\citeA{beckers68,beckers72,bray.et74,athay76}).  From the later 1970s,
numerical methods were becoming more practical due to improvements in computing power, resulting in
detailed numerical models of spicules (reviewed by \citeA{sterling00}).  

A series of models assumed that
hydrodynamic waves and/or shocks channeled by the magnetic field generated spicules 
(e.g., \citeA{hollweg82,suematsu.et82,shibata.et82,sterling.et88,blake.et85,cheng92}).   

A subset of these models (e.g., \citeA{hollweg.et82,hollweg82,sterling.et88,cheng92,sterling.et90}) 
relied on {\it rebound shocks} to create the spicule.  In the 1D simulations these shocks result when 
a pulse of $\sim$1~\kms\ (say, induced by granule motions of at the base of a vertical magnetic flux tube.  
The pulse propagates upward and non-linearly evolves into a series of shocks (the rebound shocks, formed
when some of the material raised by a shock falls and rebounds off of the denser material below).  
These shocks then lift the transition region, and also shock-heat the chromospheric material below the 
transition region.  This raised and heated chromospheric material was proposed to be the spicule. The
simulated non-steady rise of the spicule top could match early reports that some spicules appear
to rise in jerks \cite{bray.et74}.
There were however several difficulties with this model, such as whether there was a proper accounting
for the radiative losses.  Also, it was not certain whether spicules showed the non-steady upward
motion predicted as the rebound shocks successively strike the bottom side of the transition region
at (roughly) the acoustic cutoff frequency of the chromosphere.  (More recent 
observations from \hinode, e.g.\ \citeA{depontieu.et07a} do 
{\it not} show spicules to have such drastic non-ballistic motions, and so the earlier reports of 
jerky trajectories may have been due to terrestrial atmospheric effects.) 

Another series of these models \cite{suematsu.et82,shibata.et82,shibata_suematsu82} has 
a similar initial atmospheric setup, but relies on an initial pulse that is stronger than in the 
rebound shock model.  Among the predictions for a spicule resulting from 
this process was that a base brightening would be expected at the spicule start time, due to 
the large initial energy input.  Observations in H$\alpha$ of \citeA{suematsu.et95} however 
did not find such base brightenings at the expected times.

Other models \cite{shibata.et82,sterling.et91,sterling.et93} considered whether an initial energy deposition in
the chromosphere (instead of the photosphere as in many other simulations) could drive spicules.
This resulted in model spicules with rise trajectories closer to ballistic (and hence similar to 
later observations), but still predicted base brightenings that do not appear to match the observations
(e.g., \citeA{suematsu.et95}).  A later study \cite{guerreiro.et13} also shows that this 
idea has difficulties matching observations.

Yet another focus was models based on \al\ waves entering the base of a vertical flux tube,
that non-linearly couple to other MHD modes to generate the spicules 
\cite{hollweg.et82,mariska.et85,kudoh.et99}.  More-recent versions of this idea are continuing
today (e.g., \citeA{cranmer.et15}; and \citeA{iijima.et17}, which is discussed in \S\ref{subsec-recent spicule theories}).

A conclusion of the \citeA{sterling00} review was that the resolution to the spicule problem
was still outstanding, with progress required on the observational front to guide theoretical
studies.  Some such progress has come over the subsequent two decades.

\subsection{More-Recent Spicule Observations}
\label{subsec-recent spicules}

With its launch in 2006, the \hinode\ spacecraft \cite{kosugi.et07} provided the first long-term 
high-resolution observations of the solar chromosphere from space, with its 50~cm Solar Optical Telescope 
(SOT; \citeA{tsuneta.et08}).  

\subsubsection{``Type-I, ``Type-II,'' and ``Classical'' Spicules}
\label{subsubsec-spicule types}

SOT limb views revealed spicules with characteristics quite different from those discussed
previously.  From observations at the limb in \caii\ with \hinode/SOT, along with earlier work reported in 
\citeA{depontieu.et04}, \citeA{depontieu.et07a} concluded that there were two different spicule populations,
which they called type~I and type~II spicules.

Type~II spicules are prevalent in coronal holes and quiet Sun; they have relatively high 
velocities (30---150~\kms), lifetimes of 10---150~s, and usually fade from view without falling 
back to the surface (a later study by \citeA{skogsrud.et15} 
shows that often a faint trace of the falling type~II spicules can be seen with sufficient image-intensity 
enhancement).  Type~I spicules seem to be most prevalent in and around active regions.  They have 
upward velocities of 15---40~\kms, lifetimes of 3---10~min, and are seen to rise and fall back to the 
surface. (Some of these values are from an updated summary in \citeA{pereira.et14}.)   

Even prior to this usage of the term ``Type~I spicules,'' small jet-like features were identified in plage regions with
high-resolution ground-based images, sometimes observed in conjunction with with UV images and given various designations, 
including ``dynamic fibrils'' (e.g., \citeA{hansteen.et06}).  Extensive observations (\citeA{berger.et99,depontieu.et99,depontieu.et03}, 
and both 2D (e.g., \citeA{hansteen.et06}) and 3D MHD simulations 
(e.g., \citeA{martinez-sykora.et09,heggland.et11}) support that these features 
(along with at least some quiet Sun mottles, \citeA{rouppe.et07}) are driven by acoustic waves.  Most of
these show oscillations close to the acoustic cutoff frequency (\citeA{hollweg82}), with additional input from $p$-mode
waves on sufficiently inclined flux tubes (\citeA{suematsu90,depontieu.et04}).  These studies provided insight into 
investigations of the \hinode-era ``type~I'' spicules.

This type~I/type~II designation for spicules however has not been universally accepted, with \citeA{zhang.et12} maintaining
that there is no fundamental difference between spicules in the different regions.  \citeA{pereira.et12} however
contest the \citeA{zhang.et12} findings, and present counterarguments.  

(Another possible point of 
confusion with the terms ``type~I'' and ``type-II'' spicules, is that the same terminology was 
used earlier by \citeA{beckers68}, where his meaning is completely {\it different} from 
type~I and type~II as used by De~Pontieu~et~al! Beckers' type-I/type-II refer to spicules with Ca line widths
that are respectively broader and less broad; see \S~3.3.1.4 in \citeA{beckers68} for details.  In this work, 
and in most of the discussions in the literature on spicules in recent years, the terms refer to 
the spicules as defined by \citeA{depontieu.et07a}.)


Most of the earlier ground-based observations were in H$\alpha$, and therefore there are inherent differences from the
spicules observed in \caii\ by SOT\@.  And moreover, the specific \caii\ filter of SOT is different from that used in
ground-based \caii\ observations.  Therefore, it is not clear -- at least not to this author -- that SOT sees the same
component of spicules as that described in the historical literature on spicules from ground-based 
observations.  (And indeed, there could be differences in what was observed and described by different ground-based
observations.)  That is, the SOT \caii\ passband may be seeing limited aspects of the spicules described earlier.
In order to be clear when discussing the newer observations in terms of the older ground-based observations (and newer
observations taken in the manner of the older observations from the ground), we call the spicules seen in ground-based 
observations from prior to approximately the year 2000 (and hence, including
the Beckers' era spicules) ``classical spicules.''  This terminology was used in 
\citeA{sterling.et10a} and also by \citeA{pereira.et13}.

Assuming for the moment that there are two types of spicules (that is, assuming for the moment that 
\citeA{depontieu.et07a} is correct in saying that there are two types
of spicules, contrary to \citeA{zhang.et12}), we might ask which of the two 
types corresponds to the spicules observed over the first 100 years of spicule studies?  I maintain that 
the classical spicules are type~II spicules, for the following reason.

Earlier (i.e.\ classical) investigations are either unclear on whether spicules exist over plage regions, or they say that they 
are absent over plages \cite{michard74,zirin74,zirin88}.
Moreover, the tallest, most pronounced spicules studied during that earlier
period were those in coronal holes.  (\citeA{gaizauskas84} does report 
spicules at the {\it borders}
of active regions, but does not talk about spicules over any active region itself.) It is true that 
most of those observations were with various resolutions, time cadences, and likely varying 
quality of atmospheric seeing, and therefore the physical 
properties derived from those earlier periods would indeed be subject to question.  But the location on the
Sun in which they occurred would be quite reliable.  The classical spicules were generally observed
in quiet Sun and coronal holes.  Because that is precisely where the type~II spicules are positively 
observed, {\it most classical spicules must correspond to the type~II spicules}.  \citeA{pereira.et13} 
also present arguments that type~II spicules look like classical
spicules in many aspects, if the SOT images are degraded in time and cadence to mock the ground-based
circumstances. For example, in the original data the mean maximum velocity of their type II spicules was
60.0~\kms.  In their degraded data, this was reduced to 25.8~\kms, which is close to the reported
values for classical spicules.  This procedure however did not show as high a percentage of spicules 
falling back to the surface as quoted in the classical-spicule literature; although the classical results are
mainly derived from H$\alpha$ rather than \caii\ as used by SOT, and so the comparison is tricky in many 
regards.

The type~I spicules on
the other hand are reported to be seen primarily in plage regions.  Thus the type~I spicules likely were 
rarely if ever observed during the classical period.
Indeed, using \hinode/SOT images, \citeA{anan.et10} found spicules in a plage region to 
average only $\sim$1000\,km in length, which is substantially shorter than those found in quieter regions. 
Therefore, type~I spicules were probably not reported in classical observations due to their
shorter extent in plage regions rendering them more difficult to detect.  
(\citeA{shibata.et82} and \citeA{sterling.et88} [via their lower initial-TR-height case]
discuss the question of the near-absent appearance of spicules in plage regions in terms of the
wave pulse models.) 



So in summary: Classical spicules are those discussed by \citeA{beckers68} and \citeA{beckers72},
meaning seen from the ground mostly in H$\alpha$ (with the inherent limitations of ground-based observations), 
mainly in CHs and
QS regions, with upwards velocities of 25~\kms\ and other properties as described above.   Type~I and type~II 
spicules are as defined by \citeA{depontieu.et07a}: type~IIs (observed in SOT/\caii) dominate in CHs and QS, 
have upward velocities of
$\sim$30---150~\kms, generally fade without falling, etc., as described above.  Type~Is occur in or near ARs, 
have upward velocities of $\sim$15---40~\kms, show both up and down motions, etc., as described above.

And: If there are two types of spicules, there is supporting evidence that the type~IIs correspond most closely to
the classical spicules.  The average quoted speeds and lifetimes of type~I spicules is closer to the classical average
values than are the type~II spicules, but as just described the match between the classical and type~II average values
are closer when the quality of the images of the type~II spicules is degraded as presented by \citeA{pereira.et13}.

(Several works equate type~I spicules with classical spicules, but as I have just argued, this association is
not correct; or at least, it is unclear and confusing.
Some papers that make this association are \citeA{scullion.et11},
\citeA{tsiropoula.et12}, \citeA{klimchuk12}, and
\citeA{klimchuk.et14}; in none of these cases does this association have consequences for the main themes of
those works.)


\subsubsection{Other Aspects of Newer Spicule Observations}
\label{subsubsec-other new observations}

We briefly touch on a couple more aspects of spicule observations, with an emphasis on insights gained
from newer studies.

Because classical observations had long indicated that some spicules observed in chromospheric spectral lines fade,
it was long suspected that spicules are heated as they rise (e.g., \citeA{pneuman.et78,sterling.et84,sterling98}).  
Such evolution from cooler to warmer spicules has received
strong support in recent years by combining observations from SOT with those from AIA \cite{depontieu.et11} and
and IRIS \cite{depontieu.et14b,pereira.et14,skogsrud.et15}, showing a progression of emission from cooler 
(chromospheric) spectral lines to hotter (UV and EUV) lines.  These observations also showed definitively that at least
some chromospheric and transition-region spicules have a one-to-one correspondence with each other.
Several other studies have also considered the question of whether spicules have counterparts at transition region 
and coronal temperatures (e.g., \citeA{tian.et14,samanta.et15,jiao.et15}).

There also has been suspicion from classical observations that spicules display twisting motions, due to
some of the observed spectral lines appearing tilted when the spectral slit is placed normal to the length of
the extended spicules, consistent with the two sides of the spicule moving in opposite directions \cite{beckers68,livshits67,pasachoff.et68,rompolt75}.
Newer observations now provide supporting evidence that as least some spicules twist \cite{suematsu.et08,depontieu.et12}.  
Additionally, some UV spicules similarly likely show twist, \cite{curdt.et12,depontieu.et14a}. \citeA{zaqarashvili.et09}
review oscillations and waves in spicules.

Spicules are essentially invisible in SOT \caii\ on-disk observations \cite{beck.et13},
unless extremely close to the limb \cite{sterling.et10b}.  Improvements in ground-based techniques 
however reveal new features that are suspected of being counterparts to some of the limb-observed high-speed
spicules; these include so-called ``straws,'' \cite{rutten07}, ``rapid redshifted excursions,'' (RREs) and ``rapid blueshifted 
excursions'' (RBEs)   (e.g., \citeA{langangen.et08,rouppe.et09,sekse.et12,sekse.et13a,sekse.et13b}). This
is a new and involved topic that we do not discuss further here.

\subsection{More-Recent Spicule Models}
\label{subsec-recent spicule theories}

Since the review of earlier spicule models by \citeA{sterling00}, among the ideas for spicule 
generation that has
garnered much attention is that they result from the leakage of photospheric $p$-mode oscillations into
the upper atmosphere. This concept differed from some of the earlier pulse models 
(e.g., \citeA{hollweg82,suematsu.et82}) in that those earlier models only considered pulses of 
frequency below the chromospheric acoustic cutoff frequency, which corresponds to a period of 
about 220~s in the chromosphere.  This means that $p$-mode waves, 
of periods $\sim$300~s, would be expected to be evanescent in the chromosphere.
\citeA{depontieu.et04} showed that using an input spectrum of $p$-modes
in a numerical simulation could reproduce model spicules with characteristics that closely matched
spicules observed in an active region.  It turns out that - in the simulation - the initially evanescent 
chromospheric waves driven by the $p$-mode oscillations can propagate far enough into the
chromosphere to reach a layer where they again become propagating, and $p$-mode oscillations on a steeply 
enough inclined-from-vertical
flux tube propagate instead of being evanescent.  These ``leaked'' $p$-mode waves result in the spicules in the
model. \citeA{suematsu90} did a preliminary investigation of $p$-mode waves on inclined flux tubes, but 
it received increased attention, 
in particular since  
\citeA{depontieu.et07a} argued that many of the spicules they refer to as ``type~I" are produced by such a
mechanism.

Observations of dynamic fibrils (\citeA{hansteen.et06,depontieu.et07b}; also see
\citeA{marsh76}), and at least some quiet-Sun mottles \cite{rouppe.et07}, 
show oscillations both near the acoustic cutoff frequency and the $p$-mode frequency, consistent with wave
models.  \citeA{depontieu.et07a} provide evidence that the spicules they refer to as
type-II behave differently, and are thus likely to be driven by a different mechanism.  


Modeling features with characteristics of type~II spicules has been more challenging.  For example, simulations of 
\citeA{martinez-sykora.et11} and \citeA{martinez-sykora.et13} using a 3D MHD  code produce model spicules with the reported type-II 
properties, but not with a frequency of occurrence needed to match the true Sun; spicules are rare in that model but common on the Sun.  

Much more promising are simulations of \citeA{martinez-sykora.et17}, who find that, with the addition of ambipolar diffusion processes
they can produce copious type~II spicules in a model solar atmosphere.  Their simulations suggest that spicules result when ``magnetic
tension is amplified and transported upward through interactions between ions and neutrals or ambipolar diffusion."

Other recent spicule models include that of \citeA{iijima.et17}, which drives spicules
via \al\ twists at the spicule base, following on to work of \citeA{hollweg.et82} and \citeA{kudoh.et99}. Their resulting
simulated spicules look realistic, and so their work supports that this process may explain some spicules.  In order to reproduce
spicules of large enough heights however, their simulations appear to require motions near the photosphere of 5---10 \kms\ 
(see Fig.~7 of \citeA{iijima.et17}).  Historically, typical reported
photospheric motions due to granules are only $\sim$1~\kms, e.g.\ \citeA{foukal13}, and so an interesting 
question is whether improved high-resolution
observations of granules, intergranular flows, and/or intergranular magnetic elements, might show the required velocities.

In a different approach, \citeA{judge.et11} suggest that type~II spicules might be warped current sheets that only appear
to be long, thin, fast-moving features due to a kind-of illusion when the sheets are viewed in superposition.  \al\ fluctuations
in the sheets driven from below would produce the apparent motions.  They argue that this could explain the sudden appearance of
some spicules along their full length; an alternative explanation for this sudden appearance is that it results from torsional 
motions of spicules suddenly becoming Doppler shifted into the observed wave band (\citeA{sekse.et13a,sekse.et13b}).

\citeA{sterling.et16a} present arguments that spicules work like miniature versions of 
the larger-scale coronal jets, building on an idea presented earlier by \citeA{moore.et77} (also see \citeA{rabin.et80}, 
\citeA{moore90}, and \citeA{moore.et11}).  
We will revisit this suggestion in Section~\ref{subsec-small jets}, after 
discussing jets in the next Section.

The field of spicule research is again extremely rich, and many other studies of them not discussed here have 
been carried out in recent years.  Moreover, active observational and theoretical
studies based on past and current data are still underway.  Continued improvements in 
observational capabilities (e.g., DKIST, as well as new telescopes at Big Bear Solar Observatory and 
in China) should allow for new large strides in our understanding of spicules.

\section{Coronal Jets}
\label{sec-jets}

One of the more striking early features observed with the soft X-ray telescope (SXT; \citeA{tsuneta.et91}) on the 
\yohkoh\ spacecraft (launched in 1991) were dynamic jets shooting out from the lower atmosphere into 
the corona. These SXT observations motivated jet studies over the ensuing two decades.  Similar to
the situation with spicules however, there have been two eras of jet studies, albeit on a much shorter
time frame than the century-plus of spicule studies. (Although surges, at least some of which are cool 
components of coronal jets, have been long observed from the ground.)  After the initial \yohkoh-era work, new observations 
with the X-ray telescope (XRT) on \hinode\ (launched in 2006), and with instruments on the Solar Dynamics
 Observatory (\sdo, launched in 2010), spearheaded
new understanding of jets.  On \sdo\ the key instruments were its Atmospheric Imaging Array 
(AIA; \citeA{lemen.et12}), and its Helioseismic and Magnetic Imager \cite{scherrer.et12}.  
The presentation below will cover both of these periods. We will however give 
greater emphasis to developments in the later era, as the earlier period was covered much greater
detail in the extensive review of \citeA{raouafi.et16}.

\subsection{Earlier Jet Studies: Observations}
\label{subsec-early jet observations}

There were indications of the nature of coronal jets prior to SXT \cite{raouafi.et16}, but they were only fully 
appreciated with the new long-term and nearly continuous high (for the time) cadence and resolution capabilities 
of SXT \cite{shibata.et92,strong.et92}.  \citeA{shimojo.et96} and \citeA{shimojo.et98} examined the properties of 
the jets in detail, finding values that generally covered wide ranges, for example, lengths ranging from 
$\sim$10$^4$---$10^5$~km, widths of 5000---10$^5$~km, velocities from tens of \kms\ to over 1000~\kms\ (averaging 200~\kms),
and lifetimes covered from minutes to hours; \citeA{shimojo.et00} found jet temperatures to range over
3---8~MK, with an average of 5.6~MK\@.  In some cases, the wide ranges are a consequence of the variable cadence
and spatial resolution used in the SXT observations.  For example, while SXT was capable of 
high cadences of a few seconds, it also had large gaps in coverage, due to the satellite 
day-night cycle and other factors.  Jets in ARs can sometimes occur in rapid succession \cite{sterling.et16b}, 
but these might be inferred as a single jet without the benefit of continuous coverage at sufficient cadence 
($\ltsim 1$~m).   If, for example, the footpoints of those multiple 
jets were only slightly offset from each other, low-cadence observations might conclude that there is one exceptionally 
broad jet occurring from that area, while higher cadences could reveal two or more jets.

A common aspect of essentially all jets is that they have a bright point, or, in more detail, a small bright loop
or arcade of loops, at 
the base of the spire that makes up the extended part of the jet.  In many jets this 
bright point is located asymmetrically off to one side of the spire, rather than directly beneath the spire.  For 
convenience we will call this brighting the jet-base bright point (JBP) (cf.~\citeA{sterling.et15}). The JBP is
frequently more prominent in X-rays images than in, say, EUV, indicating that it is substantially hotter than 
other parts of the jet and the nearby corona.

Jets observed with SXT are, of course, X-ray jets.  Moreover, \citeA{shimojo.et96} found that most SXT jets (68\%) 
occurred near ARs.  As we will see shortly (Section~\ref{subsec-late jet observations}), coronal jets (not 
all of them necessarily detectable 
in X-rays) are also plentiful in coronal holes and quiet Sun.  The XRT bias toward the AR jets is likely due to 
the temperature response of XRT \cite{tsuneta.et91}, which tends to see hotter plasmas than does, for 
example, XRT on \hinode\ (Section~\ref{subsec-late jet observations} below).

Following the initial X-ray observations of coronal jets by \yohkoh, they have been observed in EUV by several 
different instruments.  Several instruments on the SOHO (\soho, launched in 1995) satellite contributed to jet 
studies.  The \soho/Extreme Ultraviolet Imaging Telescope (EIT) took images at four different 
EUV wavelengths: 171, 193, 284, and 304~\AA\@.  Its cadence was 12~min, which limited somewhat 
its ability to study the evolution of jets, since this is comparable to the lifetime of many jets 
\cite{savcheva.et07}.




\subsection{Earlier Jet Studies: Theories}
\label{subsec-early jet theories}

To create brightenings in the corona of the intensity of jets requires substantial energy, and from early on it
was suspected that they are powered by magnetic reconnection.  \citeA{shibata.et92} recognized the 
JBP as a clue to the cause of the jet.  They proposed that reconnection between a newly emerging magnetic
bipole and pre-existing ambient far-reaching magnetic field was the source of the jet.

\begin{figure}[ht!]
\hspace*{0.2cm}\includegraphics[angle=0,scale=0.65]{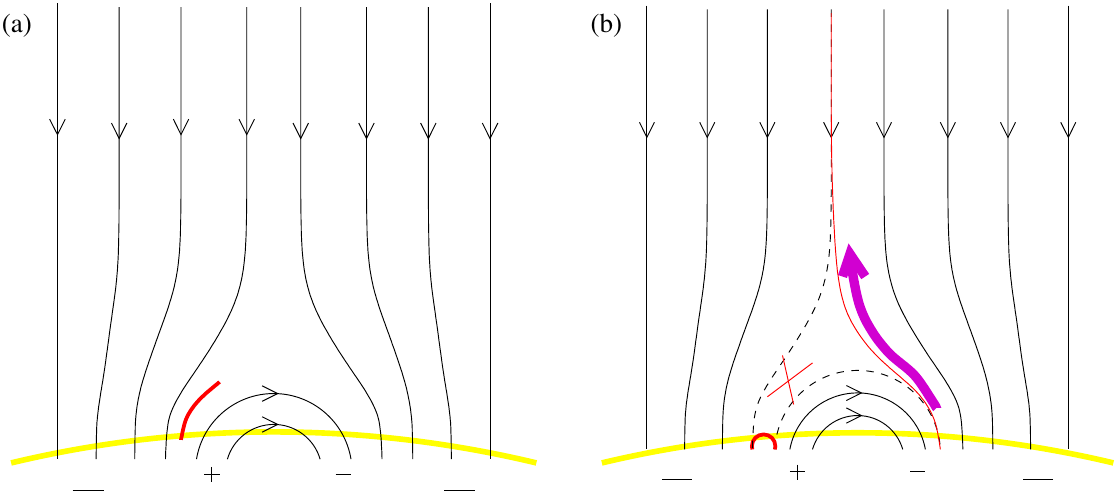}\vspace{0.0cm}
\caption{Schematic representation of the emerging flux model for coronal jets.  \citeA{shibata.et92} suggested that a 
jet results when 
(a) bipolar field emerges from beneath the solar surface, forming a current sheet where the field lines are oppositely
directed.  (b)  Continued emergence results in reconnection at the current sheet (red X), with one reconnection product 
being a small loop (red semi-circle), which was proposed to represent the jet bright point (JBP); and the other being a 
new open field line (red line), along which the jet spire flows outward (purple arrow).  Dotted lines represent the field
that reconnected to give the JBP and the new open field.  This schematic appeared in \citeA{sterling.et15}. \label{fig3}}
\end{figure}

Figure~3 shows a schematic of this proposed emerging-flux mechanism.  Here, the background field is
of a single polarity (negative in this schematic), as in, say, a coronal hole region.  When a bipole emerges into 
this region, clearly the polarity of one of the bipole's two polarities will match that of the background, 
and the other pole will be of the opposing
polarity.  As emergence continues, a current sheet develops at the interface between the bipole and
the open field, and eventually magnetic reconnection can occur along this sheet.  A result would be 
two post-reconnection products, one of these would be a new open field line, along which reconnection-heated 
jet spire material could flow.  A second product would be a new loop that is smaller than the emerging-bipole
loops (at least in the geometry of Figure~3).  \citeA{shibata.et92} suggested that the new small loop
was the bright point (later called the JBP).  Moreover, in addition to the flows induced by reconnection 
heating, a slingshot effect would assist in propelling the spire material outward, as the open-field
reconnection product snaps away from the reconnection point due to magnetic tension.

At the time it was suspected that surges in ARs resulted from the same type of emerging-flux mechanism 
(\citeA{heyvaerts.et77}, and references therein).  And in fact subsequent observations showed that surge-like
cool-material ejections can also accompany X-ray jets \cite{canfield.et96}.

\citeA{yokoyama.et95} and \citeA{yokoyama.et96} performed numerical simulations of the emerging-flux idea, resulting in
features that resembled observed jets.  They also found that, by assuming that the emerging flux loop runs
into an overlying coronal field that is horizontal rather than vertical or oblique, they could also reproduce 
features that look like ``two-sided loop jets,'' which had also been observed 
\cite{shibata.et94,yokoyama.et95,yokoyama.et96}

Over the next 20 years, there were several simulations that included refinements to this
basic idea (e.g., \citeA{yokoyama.et01,miyagoshi.et04,nishizuka.et08,moreno-insertis.et13,archontis.et13,fang.et14}).
(See the \citeA{raouafi.et16} review for a complete discussion of these earlier models.)

A different set of numerical simulations were a variation on this emerging-flux idea.  Motivated, at least in part,
by the fact that several jets and jet-like features are observed to spin \S\ref{subsec-magnetic causes} 
\cite{pariat.et09,pariat.et10,pariat.et15}.  These simulations initiate the jet by adding an \al ic shear 
to the base of the jet, leading to an outward eruption and jet flows.

\subsection{Later Jet Studies: Observations}
\label{subsec-late jet observations}

A new era of jet studies began with the \hinode\ and \sdo\ satellites.  \hinode\ carried the X-ray telescope 
(XRT; \citeA{golub.et07}), the followup to \yohkoh's SXT\@.  Because of its Sun-synchronous orbit, \hinode\ could carry
out observations for extended periods of time without day-night cycle interruptions.  XRT has increased spatial
resolution over SXT ($1''$ pixels compared to SXT's $2''.5$ pixels).  Its spectral response 
\cite{narukage.et11} extended to ``softer'' (less energetic) X-rays than SXT, in particular with its
Al-poly, Al-mesh, and C-poly filters, so that it could effectively detect plasmas in the $\sim$1---2~MK 
range.  Such temperatures were too cool for easy detection by SXT (under comparatively quiet conditions SXT 
could detect plasmas of $\sim$1.5~MK; \citeA{sterling99}).  

A set of early dedicated observations by XRT looking at the north polar region revealed that coronal jets are
prominent and frequent in solar polar coronal holes \cite{cirtain.et07}, occurring at an estimated rate of 60/day
in the two polar coronal holes \cite{savcheva.et07}.  These polar coronal hole (PCH) jets could 
reach $\sim$50{,}000~km, and
had lifetimes of about 10 minutes.  There seem to have been few studies of polar coronal hole
regions with SXT, but one such study (with its AlMgMn ``thick filter,'' sensitive to temperatures near 3---5~MK) 
did possibly detect jet brightenings \cite{koutchmy.et97}.

Studies of the temperatures of coronal hole jets provide an explanation for this.  \citeA{pucci.et13} used
the ratio of intensities in different XRT filters for one jet, and \stereo\ EUV filter intensity ratios
for a second jet, to determine their temperatures using the filter-ratio 
method.  Similarly, \citeA{paraschiv.et15} used the intensity ratios among 
different XRT filters to determine the temperatures of 18 polar coronal hole jets.
Both investigations gave temperatures in the range 1.5---2.0\,MK\@. These (comparatively) low 
temperatures could, at least in part, 
explain the paucity of reported SXR jets in polar coronal holes, and also the comparatively low SXT 
detection of jets outside of ARs (with 68\% occurring near ARs are noted above) from \citeA{shimojo.et96}; SXT
did not have lower-temperature sensitivity needed to detect those ``cooler" (1.5---2.0\,MK) jets.  
Moreover, studies using the \hinode/EIS spectrometer yield higher temperatures, 2---3~MK, for jets occurring 
in ARs \cite{chifor.et08,matsui.et12,lee.et13,mulay.et16,mulay.et17a,mulay.et17b}; it was such higher-temperature
jets that SXT preferentially tended to detect.

\subsubsection{Standard and Blowout Jets}
\label{subsubsec-standard and blowout}

By investigating a substantial number of jets in X-rays, \citeA{moore.et10} and \citeA{moore.et13} observed that
they could be categorized by their morphology into roughly two different types.  To begin with, all jets have a
long spire extending into the corona, and a broad base region that brightens, with the spire emanating from that
base region.  In one type of jet, the basic pattern is
that the spire of the jet remained narrow throughout its life, and the bright point was the brightest part of the 
base throughout the life of the jet, and was located off to one side of the spire.  Here, ``narrow" can be defined in
terms of the size of the base; so in this type of jet, when viewed in X-rays, the spire always remains narrow 
compared to the base.  

A second pattern seen in the data was that the jet spire would start out narrow, but then the spire would broaden
so that it became as wide as the base of the jet.  Also, the brightening at the base was often not confined to 
one location on the side of the spire, but instead the entire base brightens to an intensity rivaling or exceeding
that of the initial bright point.  

\citeA{moore.et10} and \citeA{moore.et13} named the narrow and broad jets respectively 
``standard jets'' and ``blowout jets''
for the reasons we now present.  (We alert the reader however that these explanations are no longer believed by
the authors to be fully valid!  Their current view of the explanations for the these jets are presented in 
\S\ref{subsubsec-standard_blowout2} below.)

\citeA{moore.et10} argued that the jets that remained narrow in X-rays were 
produced by the emerging-flux mechanism, whereby the emerging bipole undergoes reconnection with 
ambient field as in Figure~3, and throughout the lifetime of the jet the emerging flux arcade at the base
of the jet remains intact.   \citeA{moore.et10} named these narrow jets ``standard jets,'' because 
they were thought to follow the standard picture for jets presented in \citeA{shibata.et92}.

In contrast, \citeA{moore.et10} suspected that the broad-spire jets started out in the same fashion as the 
standard jet, with the flux emerging into the corona with an ambient field, but that the emerging bipole 
becomes destabilized and erupts explosively.   In this case, early in the eruption reconnections 
would occur between the bipole 
and the ambient field, leading to the initial narrow spire just as in the standard-jet case.  After that (however
still well within the $\sim$10~min lifetime of the jet) the bipole's eruption continues, and the 
entire body of the erupting bipole would result in
brightening over the entire width of the base region, through reconnections with the ambient field.  
Reconnections in the legs of the erupting bipole would result
in brightenings similar to larger-scale flare brightenings, illuminating the entire base region.  Because in this
scenario the key element is the emerging (or emerged) bipole that erupts out and away from the solar surface, \citeA{moore.et10}
called these broader-spire jets ``blowout jets.''   They found that, for 109 X-ray jets over those two studies, 53 were 
classified as standard and 50 as blowout, with six being classified as ambiguous.  While many jets 
clearly fit into one or the other category, for several others jets the category determination -- based on whether 
a spire is broad or narrow and other morphological factors -- are likely somewhat subjective.  Nonetheless,
the nearly 50-50 split from the Moore~\etal\ studies shows that both types of jets occur frequently, at least in
PCH regions.

Complementary observations in EUV are consistent with this basic dichotomy into standard and blowout jets;
in fact the EUV observations were used as further evidence in developing this story.  EUV 304~\AA\ observations
shows erupting filament/prominence material, especially when observing at the limb.   \citeA{moore.et10} used 
EUV 304~\AA\ observations from \stereo\ of four XRT X-ray jets, and  \citeA{moore.et13} used EUV 304~\AA\ observations from
AIA to observe all of the XRT X-ray jets of that study.  More statistics were available from the
latter study using AIA\@. They found that 29 out of 32$^1$ of the blowout jets (as identified in X-rays) had 
a cool 304~\AA\ component; this supports that an emerged bipole could be erupting, as suggested in the 
blowout-jet scenario first presented in \citeA{moore.et10}.\footnote{\citeA{moore.et13} contains a typo in \S~5, 
second paragraph, where it says ``all 29 blow-out X-ray 
jets displayed a cool component.'' This should instead say: ``29 out of 32 blow-out X-ray jets displayed 
a cool component.''  This typo has no consequences for the subsequent discussions of that paper.}

Several other papers investigated blowout jets, including \citeA{hong.et17}, \citeA{chandra.et17} 
\citeA{zhu.et17}, \citeA{joshi.et17}, \citeA{shen.et17}, and \citeA{joshi.et18}.

\subsection{Later Jet Studies: Theories; the Cause of Jets Revisited}
\label{subsec-late theories}

\citeA{adams.et14} studied a jet in an on-disk coronal hole using \sdo\ AIA EUV and HMI magnetic field 
observations, with a goal of finding the flux emergence that was hypothesized to be the agent driving the jet.  
They did not, however, find any such emergence, nor were there data consistent with the jet being caused
by a bipole that had recently emerged being the source of the jet in the manner pictured in Figure~3.
Instead, the source of the jet appeared to be a small-scale filament that erupted to form the jet spire, and
the bright point that occurred at the side of the jet appeared to be a scaled-down version of a flare that
accompanies large-scale filament eruptions.  Moreover, watching the long-term evolution of the 
magnetograms showed that that jet occurred at a location where opposing polarity magnetic elements 
converged and canceled, leading to the jet.  

Thus \citeA{adams.et14} concluded that this jet resulted when magnetic
flux cancelation resulted in a small-scale filament eruption that evolved to form the jet spire, with the
bright point being the small-scale flare accompanying the small-scale filament eruption.  This  was 
quite different from what was expected based on the emerging flux idea first presented about 20 years
earlier.  

Other studies, it turns out, had seen various elements of this conclusion.  Several workers suggested that
their observed jet(s) could be due to small-scale eruptions (e.g., \citeA{nistico.et09,raouafi.et10,hong.et14}).
\citeA{shen.et12} shows very nicely an erupting small-scale filament leading to a jet.
And going back much further, while there was no identification with coronal jets, 
\citeA{moore.et77} suggested that miniature filament eruptions produced
macrospicules, and \citeA{wang.et00} found that miniature filament eruptions are common in the quiet Sun.  
Several studies also found flux cancelation to accompany jets in some events or circumstances 
\cite{shen.et12,innes.et13,hong.et14,young.et14a,young.et14b}.

Other instruments contributing to jet studies from this period include \soho/CDS, SUMER, and UVCS, and 
also the EUV \trace\ telescope.  We refer the reader to \citeA{raouafi.et16} for detailed discussions of
jet results from these instruments.

\subsubsection{Minifilament Eruption Model for Coronal Jets}
\label{subsubsec-minifilament eruption model}

\begin{figure}[ht!]
\hspace*{-1.0cm}\includegraphics[angle=270,scale=0.65]{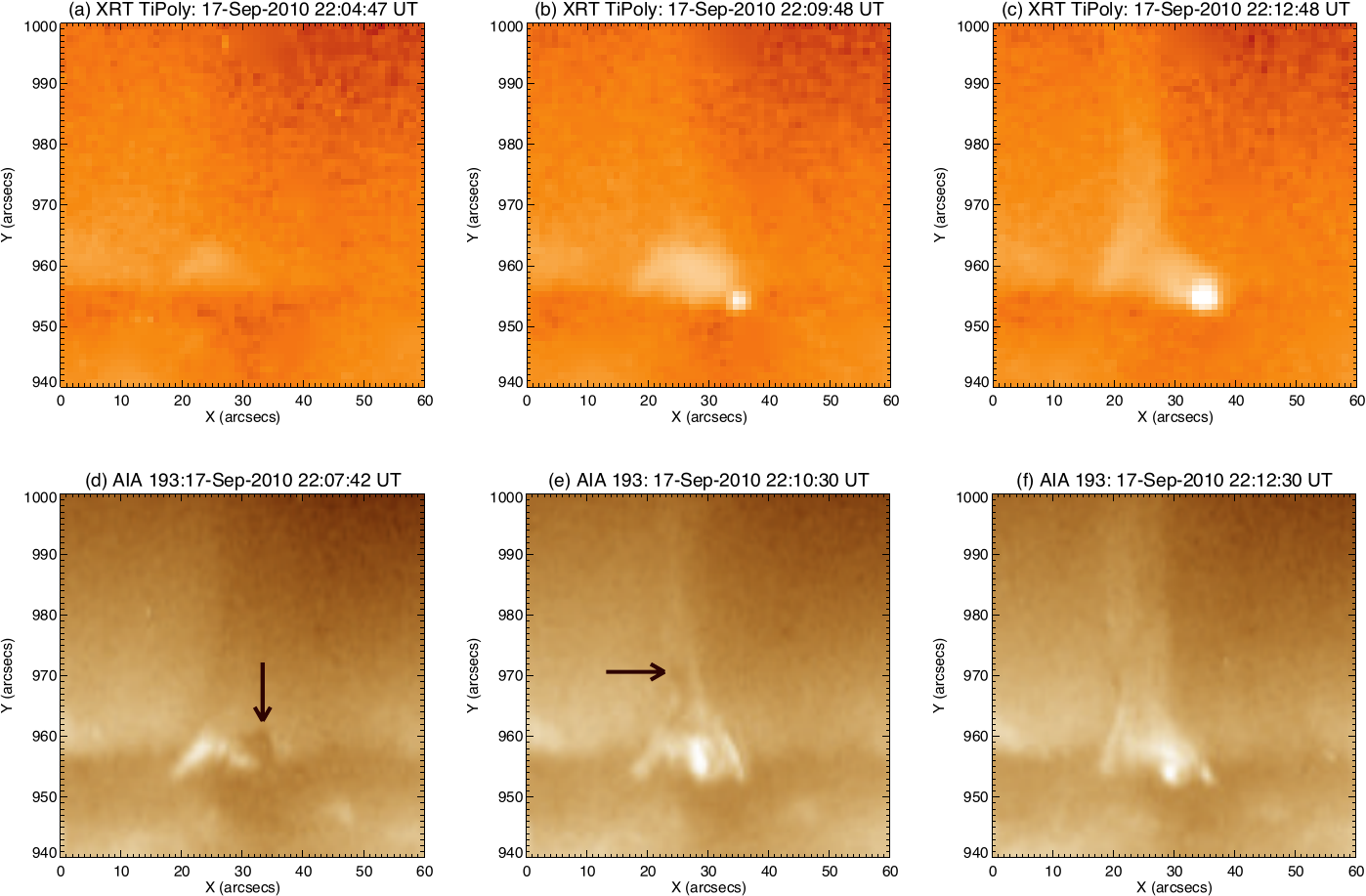}\vspace{0.0cm}
\caption{(After \citeA{sterling.et15}.) Observed X-ray jet near at the solar limb in (a) X-rays from XRT, and (b) in 193~\AA\ EUV from
AIA on \sdo\@.  EUV frequently shows dark (presumably cool) absorbing filament material erupting and traveling out along the spire, 
and the JBP appearing at the location from which the ``minifilament" erupted.  Here, the JBP is visible as the bright point-like
features in (b), that grows larger and brighter in (c).  The arrow in (d) shows the minifilament emanating from the location where the JBP
forms; in (e), the erupting minifilament is moving out along the jet spire. \label{fig4}}
\end{figure}

\citeA{sterling.et15} used the data set of \citeA{moore.et13} of near-limb PCH X-ray jets, and examined 20 of these 
jets in all of the AIA channels.  (\citeA{moore.et13} had inspected
AIA~304~\AA\ data, but not that of other channels.)  They found that {\it all} of the jets were consistent with 
having resulted from 
eruptions of minifilaments, with the bright point on one side of the jet's base (now called a JBP, for ``jet bright 
point" or ``jet-base bright point") being a miniature flare accompanying the minifilament's eruption.  Figure~4
shows one of the 20 jets \citeA{sterling.et15} found to result from minifilament eruptions; earlier studies had
shown minifilaments erupting to form jets in studies of limited numbers of events (Fig.~5). \citeA{sterling.et15} 
proposed that essentially all jets were formed by this process.  Figure~5 shows an example of a minifilament
eruption producing a jet.

\begin{figure}[ht!]
\hspace*{-2.0cm}\includegraphics[angle=0,scale=1.0]{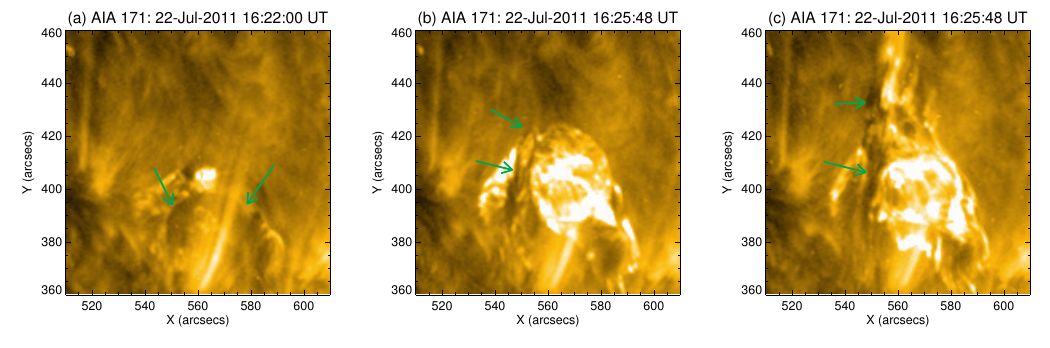}\vspace{0.0cm}
\caption{Minifilament observed erupting in \sdo/AIA 171~\AA, where the arrows in the panels point to
the minifilament, which shows up in absorption. (a) The minifilament is starting to expand upward,
while accompanying brightenings are still minimal.  (b) Strong brightenings are developing beneath the erupting 
filament, forming the JBP and auxiliary brightenings.  (c) As the minifilament eruption continues, the spire starts
to form above the bright base.  This event was studied extensively in \citeA{shen.et12}.   \label{fig5}}
\end{figure}

Figure~6 shows the schematic scenario \citeA{sterling.et15} presented for this minifilament eruption model; this figure is 
slightly modified from the original to reflect findings of \citeA{moore.et18}.  Figure~6a shows a double-bipole field
embedded in an ambient coronal field having a single magnetic polarity and extending approximately vertically 
from the surface.  This view is a good approximation to the expected situation in a coronal hole region, but it 
also applies to other regions of the Sun.  For example, the ambient field might be part of a coronal loop that 
is large compared to the double bipole where the requirement being that the loop is large compared to the double bipole 
shown in the figure.  (More precisely, the loop has to be substantially larger than its base, and large enough so 
that reflections from the far side of the loop are of no significant consequence to the jet's development; if these 
conditions hold, then the picture of Fig.~6 can be considered apropos without modification.) 

\begin{figure}[ht!]
\hspace*{-0.5cm}\includegraphics[angle=270,scale=0.85]{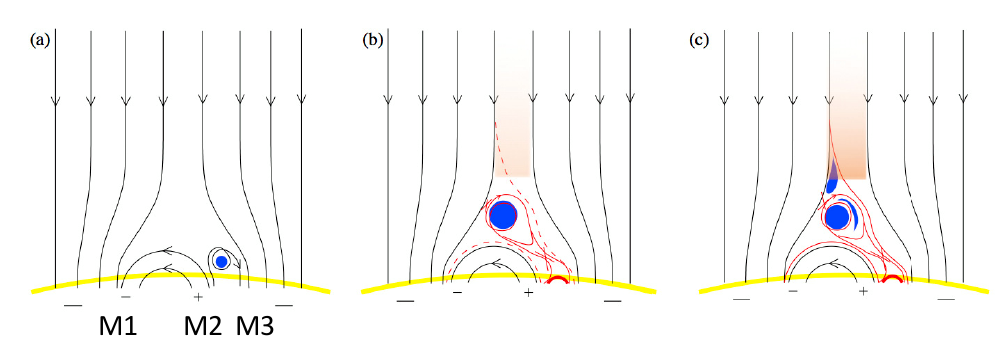}\vspace{0.0cm}
\caption{Schematic representation of the minifilament eruption model for coronal jets. \citeA{sterling.et15} proposed that jets 
result from locations where in panel (a) a miniature filament (minifilament; blue circle) is initially in a bipolar anemone region 
embedded inside of a surrounding open (or far-reaching) field.  This shows a cross-section of a 3D anemone structure, where
one side of the structure has increased stress (the smaller - more compact - loop field, rooted between the locations M2 and
M3 labeled in panel (a)) that contains the cool-material minifilament (blue circle).  (b)  Upon eruption of the minifilament, the field 
surrounding the cool minifilament reconnects with adjacent open field via {\it external reconnection} (upper red X), 
producing a new closed loop over the larger loop (between M1 and M2), and a new open field, along which the jet spire flows outward.  
Meanwhile, {\it internal reconnection} occurs among the legs of the erupting minifilament field (lower red X), producing a miniature
flare, a~la formation of a typical solar flare in the wake of larger filament eruptions as described by the 
CSHKP standard model for solar eruptions; here, the miniature flare in panels (b) and (c) (small red semicircle between 
locations M2 and M3 of (a)) is identified with the JBP\@.
(c) If the minifilament eruption progresses far enough into the open-field region, enough of the outer envelope of the
erupting minifilament field can be eroded away by the external reconnection for the cool minifilament material to flow outward along the spire as
part of the jet.  Dotted lines in panel (b) represent field soon to be made by the external reconnection.   This is a modified version of a
schematic in \citeA{sterling.et15}. \label{fig6}}
\end{figure}

In 2D cross-section, Figure~6 shows a double bipole embedded in the base of
the ambient field.    This setup would naturally occur on the Sun if an opposite-polarity field 
were embedded in the 
base of the unipolar ambient field.  In 3D that opposite polarity field would bloom out into an anemone-type
configuration \cite{shibata.et07}, so that opposite-polarity flux connects to the surrounding ambient flux;
Figure~6 shows a 2D cross-sectional cut of that 3D anemone structure.  In this cross-sectional view, 
one side of the bipole is more compact (the right-hand side in Fig.~6a) and more non-potential than the other, 
likely due to asymmetric photospheric flows and flux cancelation 
(\S\ref{subsec-magnetic causes}).  Magnetic shear and twist would build up
in that stressed side.  It is well known that
large-scale filaments form along polarity inversion lines in sheared filed such as the inversion line and field 
at location B in Figure~6, and
this is also where we observe miniature filaments existing prior to jets (\S\ref{subsec-magnetic causes} below).  

Figure~6b shows the next stage.  At some point (again due to processes discussed in \S\ref{subsec-magnetic causes}), 
the magnetic field enveloping the 
minifilament becomes destabilized and starts to erupt.  Prior to erupting, the field enveloping the minifilament 
was either a sheared magnetic arcade structure, or a magnetic flux rope.  In either case, at the latest 
it is a flux rope soon after the eruption starts. As it erupts outward from the surface, magnetic 
reconnections occur at two different locations.
One location is the field below the erupting minifilament, that is, the ``legs'' of the erupting minifilament field
come together below the minifilament as it erupts away from the photospheric magnetic neutral line.
This reconnection, which we call {\it internal reconnection} (because it is internal to the erupting minifilament
field) is identical to the reconnection that occurs beneath erupting large-scale filaments resulting in 
solar flares, in the standard model for solar eruptions, sometimes called the CSHKP model, after 
\citeA{carmichael64}, and \citeA{sturrock66}, \citeA{hirayama74}, and \citeA{kopp.et76} (also see, e.g., \citeA{moore.et01}).  
In this case, the result is a miniature flare, which we identify as the JBP,
and the addition of additional flux-rope field wrapping the minifilament.

The other reconnection that occurs as the minifilament flux rope erupts takes place in the region where the
flux rope field contacts the ambient field, a location that is an approximate magnetic null point.  We call
this {\it external reconnection}, as it occurs on the external side of the minifilament flux rope.  This is 
identical to the reconnection referred to as ``breakout reconnection" \cite{antiochos98,antiochos.et99}, 
or ``interchange reconnection'' \cite{crooker.et02,crooker.et12}.  This reconnection results in two 
products: one is an additional new open, vertical field,
and a second is new closed field on the larger bipole.  (\citeA{sterling.et01} 
show similar external reconnection in a large-scale eruption; compare that paper's Fig.~5 with Fig.~6 
here.)

Figure~6c shows that the jet forming as the minifilament flux rope field reconnects and mixes in with
the ambient field.  This results in formation of the spire, by both reconnection heating of the ambient 
coronal plasma at the site of the external reconnection, and by flows due to chromospheric 
``evaporation'' that occurs in standard flare models.  There also is likely a ``whipping effect''
that propels chromospheric material onto the newly opening external-reconnection field line as
it snaps to vertical following the reconnection \cite{shibata.et86}.  In this way, a jet spire 
naturally forms with the JBP off to the side of the base.

\subsubsection{Modified View of Standard and Blowout Jets}
\label{subsubsec-standard_blowout2}

Among the 20 events studied in \citeA{sterling.et15}, 14 were classified as blowout jets, five were 
standard jets, and one was ambiguous.  They concluded that, rather than via the emerging-flux picture of
Figure~3, the blowout jets instead work as described in
\S\ref{subsubsec-minifilament eruption model} and in Figure~6\@.  In this case, the erupting 
minifilament comes from the base bipolar region, as originally proposed in \citeA{moore.et10}, except
now there is no longer an assumption that the erupting bipole is an emerging bipole; \citeA{sterling.et15}
did not address directly the cause of minifilament formation and eruption, pointing out that both
flux emergence and cancelation would be candidates for triggering the eruption (although 
the observational evidence up until
that date had been favoring cancelation).  

Recalling the properties of blowout jets: They first appear as a narrow-spire standard jet, and then
evolve into a broader jet, where the spire can be as large as, or even larger than, the jet's base.  
How does this match up with the minifilament eruption model?  During the early stages of the jet 
formation (Fig.~6b), the erupting minifilament flux rope will only have external reconnection with
a few of the ambient vertical field lines; this will produce the observed narrow spire.  Also during
the minifilament's eruption, internal reconnections will form the JBP as the brightest part of the base, 
as observed during early stages of blowout jets.  Later in the jet development, it is observed that 
the spire becomes broader and the base immediately below the spire becomes brighter.  This fits with
the minifilament eruption model, as the spire becomes broader as the erupting minifilament flux rope
plows deeper into the ambient ``open" coronal field, making more and more new ``open" field lines through
external reconnection.  Moreover, the movement of the flux rope is away from
the JBP with time (at least in the simplified 2D picture of Figure~6), indicating that the spire should
grow in a direction away from the JBP\@.  That is, the spire drift should be away from the JBP,
rather than toward the JBP; this is in agreement with the trend of spire drift in observed jets \cite{savcheva.et09}.    
(In contrast, the emerging flux model predicts a spire drift toward the JBP 
with time; \citeA{sterling.et15,moreno-insertis.et13}).  These same external reconnections also 
progressively add more and more bright reconnection-product loops to the large lobe immediately 
beneath the spire, increasing its intensity in agreement with the observations.

What about the narrow-spire, so-called ``standard jets''?  For the five cases among the \citeA{sterling.et15} 
that fit this description, there was still evidence for presence of an erupting minifilament.  But for
those events the 
minifilaments were often faint, and in none of the cases was there a very obvious minifilament 
eruption completely away from the Sun (it is unclear whether there was a weak ejection from the Sun, but even
if so, they were weaker than those seen in typical blowout jets).  
Instead, the minifilament's eruption was arrested prior
to reaching the upper corona, or likely only part of the minifilament leaked out onto the vertical field.
Thus the minifilament eruptions in the standard jet cases seem to correspond to failed or partially-failed 
filament eruptions that have been observed in larger-scale filaments 
(e.g., \citeA{sterling.et11,li.et17}).

Thus for standard jets, the eruption would stop at about the stage pictured in Figure~6b, before the full filament flux
rope has a chance to escape into the upper corona (and also, observations indicate it usually stops before much
of the cool filament material reaches the spire, that is, prior to the time depicted in Fig.~6c).  In comparing 
this with the observed properties of 
standard jets, the external reconnection would form a hot spire, but the erupting minifilament
would not progress far enough into the ambient open field for much, if any, of the cool minifilament 
material to flow out along the open field.  This would explain why in the observations the spire remains
narrow, and why most of the standard jets would not show a cool component of 304~\AA\
material.  Meanwhile, the internal reconnection would form the JBP, but
the lack of extensive external reconnection would limit the intensity of the jet base immediately 
beneath the spire to modest levels, so that the JBP intensity is more likely to dominate the
base-region intensity throughout the life of the standard jet, as observed.

Therefore, in this view, both standard jets and blowout jets are the result of minifilament eruptions, 
but in the former case the eruption is likely more confined, while in the latter case the minifilament
erupts fully (``blows out").  This picture renders the name ``standard jet'' a misnomer, since these jets no longer
are thought to obey the previously-held standard emerging-flux picture (Fig.~3).  So the term ``standard jet''
should be thought of as a morphological description of jets, where the spire 
remains narrow throughout the jet's life and the brightness of the broader part of the base region 
does not rival that of the JBP\@.  For blowout jets, the morphological description of a narrow spire 
growing into a broader spire also holds, but the cause is now suggested to be the minifilament-eruption
picture of Figure~6 \cite{sterling.et15}, rather than an eruptive version of the emerging-flux picture 
of Figure~3 \cite{shibata.et92,moore.et10}.  

A note of caution: the translation of these properties of
standard and blowout jets to EUV-observed jets should be done with care, as the terms ``standard" 
and ``blowout" were originally made in the context of the X-ray appearance.  So, for example, it is 
possible that in standard jets the JBP does not stand out prominently compared to the rest of the 
jet base in some EUV channels, as it would in X-rays.  On the other hand, a clear and explosive
eruption of a minifilament into the far corona observed in EUV would be expected to have properties
of a blowout jet when observed in X-rays.

Other examples of minifilament eruptions causing jets include \citeA{hong.et16}, \citeA{zhang.et16}, 
and \citeA{hong.et17}.

\subsection{The Magnetic Causes of Jets}
\label{subsec-magnetic causes}

An understanding of the cause of jets requires observing them on the solar disk.  As mentioned in 
\S\ref{subsec-late theories}, there had been several observations of on-disk jets.   These had found either
cancelation, or emergence and cancelation occurring below jets 
\cite{shen.et12,innes.et13,hong.et14,young.et14a,young.et14b}.  Even earlier studies also saw jets occurring at 
where one pole of an emerging bipole underwent cancelation, and the jet occurring from the cancelation site
\cite{chae.et99,liu.et04}.

In retrospect however, none of these earlier studies found conclusive
evidence for jets 
occurring with emergence in the
absence of cancelation, with some of those studies concluding that cancelation was the primary 
trigger of the jet (e.g., \citeA{young.et14a,young.et14b}).  Following up on those studies, along with the work 
of \citeA{adams.et14}, using AIA and HMI
data, \citeA{panesar.et16a} looked at ten quiet Sun coronal jets, and \citeA{panesar.et18a}
looked at 13 on-disk coronal hole coronal jets.  For all of these jets, they found that they resulted from
erupting minifilaments, and that magnetic flux cancelation occurred at the minifilament location in the hours preceding 
minifilament eruption (Fig.~7).

\begin{figure}[ht!]
\hspace*{0.2cm}\includegraphics[angle=270,scale=0.50]{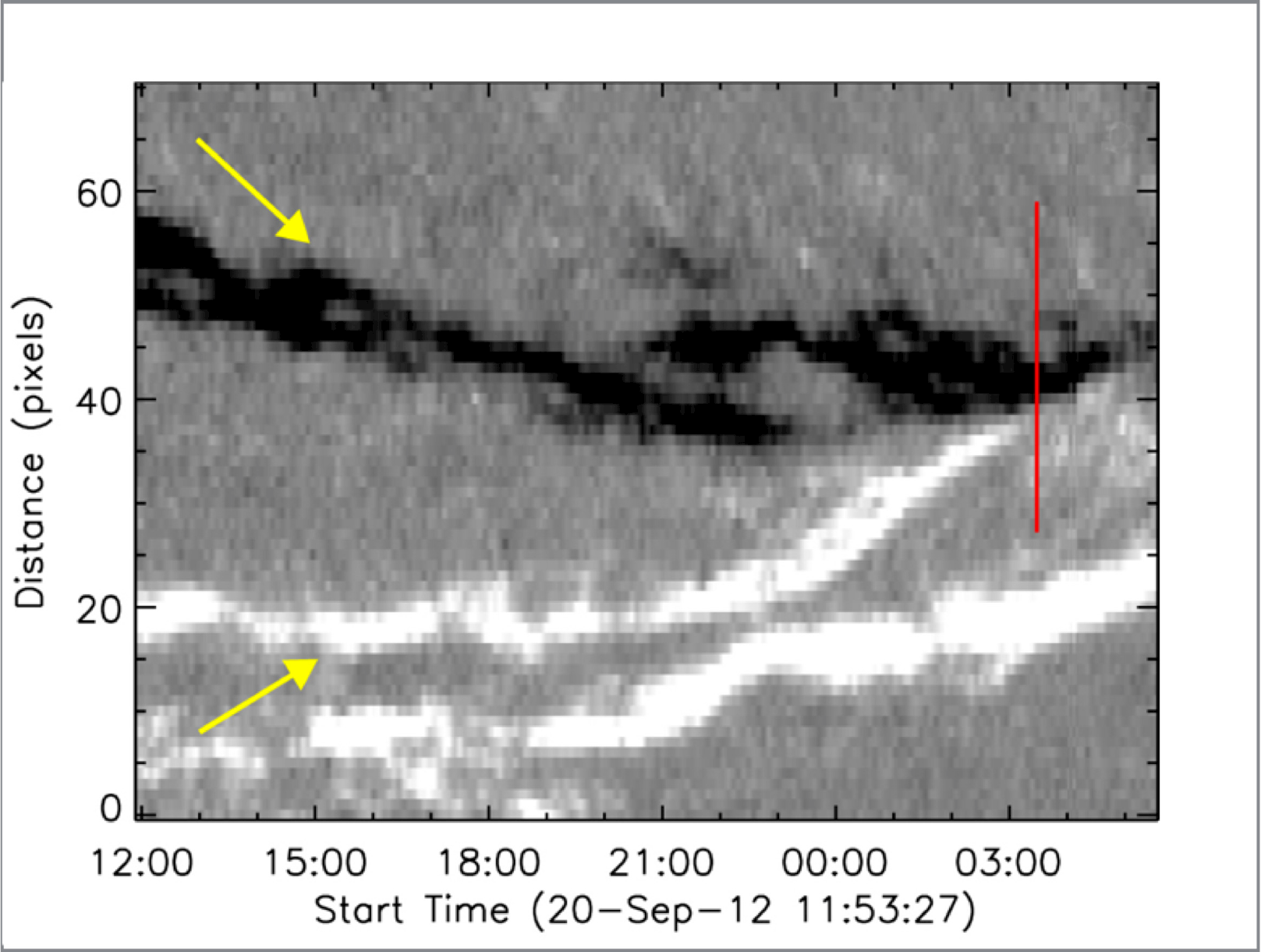}\vspace{0.0cm}
\caption{(After \citeA{panesar.et16a}.)  Plot showing magnetic evolution as a function of time of positive- (white) 
and negative- (black) polarity fields at the location where a coronal jet occurred on 2012 September~20.  
These magnetic data are from the \sdo/HMI magnetograph, which has pixels of $0''.5$.  At the 
start of the plot, the polarity elements (arrows)) are about a distance of $15''$ apart, and 
over the next 15~hr flow together, eventually converging. At the time of convergence (red line), a minifilament
eruption occurred, producing the observed coronal jet.  \citeA{panesar.et16a} discusses this event, and nine others
like it, in detail. \label{fig7}}
\end{figure}

For flux-cancelation-rate measurements for
all of the studies listed in that table, the procedure was to look at the evolution of the magnetic field in 
the region around where the cancelation occurred, over the hours leading up to the cancelation.  Among the
canceling elements, one of the polarities - the majority polarity - was usually dominate in the region, while the 
other polarity - the minority polarity - was less prevalent.  Panesar~\etal\ tracked the flux inside of only one of those
polarities, usually the minority polarity, because it was easier to isolate inside of a prescribed box as 
the cancelation occurred.  This isolation is important when determining the total integrated flux variation with 
time.  If instead the flux is not sufficiently isolated, then any flows of flux into or out of the box over time 
would confuse the results: in that case it would be very difficult to determine whether an increase or decrease 
in the measured flux value was due to 
a true emergence or cancelation on the one hand, or due to a flow into or out of the bounding box's boundary 
over time on the other hand.  More sophisticated methods also exist for tracking flux changes in 
regions \cite{green.et03}.  Flux cancels at a rate of 0.5---2.0$\times 10^{18}$~Mx~s$^{-1}$ 
and 0.9---4.0$\times 10^{18}$~Mx~s$^{-1}$ in the coronal-hole and quiet-Sun jet cases, respectively.  
(\citeA{sterling.et18} provide a table summarizing these and other 
cancelation rates.)

For the ten jets of \citeA{panesar.et16a}, \citeA{panesar.et17} examined how and when the minifilaments that 
erupted to produce the jets formed.  They found that, in all cases, the minifilaments themselves formed via
flux cancelation.  Following its creation, the minifilament was present for a period ranging from 1.5 hours to
two days prior to erupting to drive the jet.  Based on these findings then, the complete picture for the ten
jets of \citeA{panesar.et16a} is that an episode of flux cancelation forms a minifilament; then, either 
continued flux cancelation, or a second episode 
of flux cancelation, leads to eruption of the minifilament and formation of the jet via the process shown in 
Figure~6.  

In the author's view, then, several of these recent studies \cite{adams.et14,sterling.et15,panesar.et16a,panesar.et17,panesar.et18a,moore.et18} 
suggest the following picture for jets in quiet regions and coronal holes.  Photospheric flows drive together minority-
and majority-polarity flux patches.  This results in flux cancelation and a buildup of a magnetic flux rope
at the canceling neutral line.  Whatever process that is responsible for causing large-scale filaments to form
\cite{vanball.et89,parenti14} works in this situation too, resulting in formation of the minifilament
along the sheared field/flux rope. (We do not attempt to differentiate between highly-sheared 
non-potential field and a bona fide flux rope in the pre-eruption phase.)  Presence of the minority flux patch 
results in the anemone geometry 
of Figure~6a, where the flux rope and minifilament form at the location indicated in that figure as a result
of the photospheric flows and flux cancelation.  Continued cancelation, or a new episode of cancelation, then
results in destabilization of the minifilament, its eruption, and jet formation, as shown in Figures~3(b) and~(c).

Even though flux cancelation is the key agent leading to minifilament formation and eruption, this does not
rule out a role for flux emergence in the jet process.  In fact, several of the events in the Panesar~\etal\ studies,
as well as other jet studies (e.g., \citeA{chae.et99,liu.et04,liu.et11,huang.et12}) find that the jets occur only 
after one pole of the
emerging-flux bipole undergoes cancelation with neighboring field.  Thus, at least in these cases, the flux emergence
plays an auxiliary role to the jet process; but the main process apparently is still flux cancelation.

There are however also reports of jets without flux cancelation.  \citeA{kumar.et18} present an example of a jet
that originates from an equatorial coronal hole location that shows no obvious flux emergence or cancelation at 
the time of the jet. 
There are however oppositely directed motions of opposite-polarity flux before and near the time of the 
jet, and they argue that the magnetic shear
from these shearing motions is responsible for the jet, consistent with the numerical modeling of \citeA{pariat.et09}.  
Thus this may be
a jet without direct cancelation.  (Clear cancelation {\it does} occur within $\sim$20$''$ of the jet location, 
But this cancelation is away from, and -- seemingly -- not connected to, the main neutral line of the jet.)
Thus, other than shearing motion, another possibility for onset of this event are that there is an (unseen) magnetic connection
between the jet location and the location of that nearby flux cancelation.  And of course there is always the possibility 
that flux cancelation is occurring at the jet site at a level too weak to be detected by HMI (although similar statements 
could be made arguing that undetected flux emergence or shearing motions cause jets too).

\citeA{mulay.et16}, looked at 20 jets and their magnetic setting.  They concluded that ten of
the jets resulted from flux cancelation, four of them resulted from from flux emergence followed by cancelation,
and that six of them were from flux emergence regions.  A difference with these
jets from the \citeA{panesar.et16a}, \citeA{panesar.et18a}, and \citeA{kumar.et18} studies is that the 
Mulay~\etal\ events occurred in the edge of ARs, rather than in quiet Sun or CHs.
Because activity in ARs is often more rapid and complex than in quieter solar regions, often flux emergence and 
cancelation occur in close proximity to each other, complicating determination of which is the key mechanism in
the episode being studied. Similarly, \citeA{sterling.et16b} observed that emergence and cancelation were concurrently
occurring in AR jets that they examined (they also found cancelation in all but one of their events).  
\citeA{shen.et12} also found a jet near an AR to occur in a region of both emergence and
cancelation at its base. In any case, the \citeA{mulay.et16} study confirms that flux cancelation is the primary 
agent for jet production in many jets.  That study also confirms the importance for continued investigations of
the magnetic source for jets, and also (perhaps) the need for cross-comparisons and/or calibrations between 
studies carried out by different workers or research groups.

Jets frequently show clear twisting motions during their rise, with recent examples including 
\citeA{patsourakos.et08}, \citeA{moore.et15}, \citeA{innes.et16}, \citeA{joshi.et18}, \citeA{bogdanova.et18}, 
and \citeA{miao.et18,liu.et18}.  Such twisting 
could be indicative of a pre-existing
twisted field on which the jet spire travels, a twisting of the field with time, or an untwisting of a 
previously twisted field.  The detailed study of \citeA{moore.et15} of 14 jets extending from the polar 
coronal hole regions out beyond 2.2R$_{\rm Sun}$ strongly supports the latter-most explanation.  In the minifilament-eruption
model view, this is consistent with the pre-eruption flux tube holding the minifilament material being 
twisted prior to eruption.  This twist would be released then the erupting minifilament undergoes
the external reconnection with the external field, via a process described by \citeA{shibata.et86}.
Thus, with this view, the twisting would be due to an {\it unwinding} of a previously twisted flux rope.

\subsection{Active Region Jets}
\label{subsec-AR jets}

Jets in ARs would be expected to be more energetic versions of the jets in QS and CH regions.  Assuming this 
however leads to some factors that are not so easily understood.  Chief among these factors is that 
some AR jets can be very prominent in X-rays and EUV, but lack an obvious accompanying erupting 
minifilament.  So a question is: is the same physics occurring without an erupting minifilament,
or is a substantially different process responsible for the coronal jets occurring in ARs?

To begin with, the jets in ARs that we are concerned with here are strictly coronal jets, such as those
appearing in X-rays (e.g.~\citeA{shibata.et92,shimojo.et96}).  These jets appear at the edge of ARs.  
\citeA{shimojo.et96} found that about 90\% of AR jets with an X-ray bright point (by which they meant 
a point-like brightening unresolved by XRT,
and so likely the same feature \citeA{sterling.et15} call the JBP) 
occurred in the western part of the AR, and they add that such jets ``tend to appear at the western 
edge of the preceding spot.''  (Smaller ``penumbral jets,'' which occur
near sunspots, will not be discussed here; see, e.g., \citeA{katsukawa.et07},
\citeA{tiwari.et16}, \citeA{tiwari.et18}.)

Since the ground breaking work of \yohkoh, there have been many studies of jets in ARs, but often without
mention of (mini)filaments (e.g., \citeA{kim.et07,guo.et13}), while other studies did not mention
them explicitly but did discuss a connection with a surge (e.g., \citeA{zhang.et14}).  

In their study of 20 jets in EUV with AIA, \citeA{mulay.et16} found that they occurred at the periphery 
of ARs, and did not mention examples of minifilaments erupting to cause the jets.  They did point out that all of their 
jets were accompanied by H$\alpha$ surges. But while \citeA{wang.et12} found that some surges that made
circular-ribbon flares came from erupting filaments, more generally the connection between surge material 
and filaments has not been clear (e.g., \citeA{roy73}). Therefore, for many AR jets, the connection 
between the AR jets and erupting minifilaments is not as clearly obvious as it is in many quiet Sun
and coronal hole jets.

In an attempt to understand whether AR jets behave in the manner described in Figure~6, that is in
the way that quiet Sun and CH jets appear to operate, \citeA{sterling.et16b} looked at a series
of jets from a single AR, using data from EUV data from AIA and \stereo, magnetic data from HMI,
and X-ray data from the Soft X-ray Imager (SXI)\@.  They found that some of the jets (three of which they
examined in detail) showed clear minifilaments erupting to cause the jets, and that these eruptions 
displayed slow-rise and fast-rise phases that are frequently seen in typical (larger-scale) filament
eruptions.  Thus they concluded that these jets followed the pattern of Figure~6, and thus of
coronal jets in non-AR regions.  All of the jets occurred at magnetic neutral lines.  

Additionally, two of the AR jets examined in detail by \citeA{sterling.et16b} and that had clear erupting minifilaments 
also showed strong ejections in AIA 304~\AA\ images, consistent with surges and sprays.  Moreover, 
the footpoint brightenings corresponding to location M1 in Figure~6a 
appeared as semi-circular 
patterns in the images, which is similar to emission seen in a large percentage of surges
(cf.~\citeA{ohman72}).  This suggests that many or all surges might be formed as in Figure~6,
consistent with the findings for a set of surges by \citeA{wang.et12}.

Several of the AR jets of \citeA{sterling.et16b} however did {\it not} show clear minifilament eruptions.
A series of these jets developed faster than the ones that showed clear minifilament eruptions.
\citeA{sterling.et17} called such faster-developing jets ``violent jets,'' and explored several of
them in detail for a different AR, using data from AIA, XRT, \iris, and HMI\@.  They again found 
that several of the violent jets did not have prominent minifilament eruptions, but they instead 
had very thin ($\ltsim$2$''$) filament ``strands'' that erupted away from the jet region, similar to
features seen by \citeA{schmieder.et13}.  Moreover, these eruptions took place at neutral lines where
clear flux cancelation was occurring, similar to what is found in many QS and CH jets.  In 
the earlier study of \citeA{sterling.et16b}, all but one of the jets occurred where there was clear canceling flux.
In the one exception, the jet occurred at the neutral line of an emerging-flux element. (In retrospect,
we now suspect that that jet occurred where that emerging (or emerged) bipole was canceling at its own main main
neutral line during the later stages of the emerged bipole's development, similar to the larger-scale
filament eruptions in \citeA{sterling.et18}).  Table~1 of 
\citeA{sterling.et18} gives the average
rate of flux cancelation measured in these studies.  More generally, the
overall magnetic setup for all of the AR jets of both \citeA{sterling.et16b} and \citeA{sterling.et17}
was as pictured in Figure~6. These observations support that the AR jets are largely identical to
quiet Sun and coronal hole jets.

Still however, it is not clear why some AR jets are accompanied by negligible visible minifilaments.
It is possible that a magnetic flux rope is erupting from the location of the expected minifilament,
but that field merely does not contain enough cool material to appear as a cool minifilament, similar
to how large-scale eruptions can occur without an accompanying filament eruption.  One more puzzle
with many of the violent jets is that the larger magnetic lobe (between locations M1 and M2 in Figure~6a)  
adjoining the location of the minifilament eruption (or of the expected minifilament-field eruption, in the 
case the field is not marked with cool material), can far exceed the intensity from the expected-JBP location
(between locations M2 and M3 of Figure~6a); that is, in QS and CH jets, the JBP prominently stands out as
a bright part of the base, at least early in the jet development, but sometimes the JBP location intensity
remains subdued in AR jets.
\citeA{sterling.et17} speculated that this could be due to secondary eruptions triggered by eruption
of the first jet, similar to what is observed in some large-scale eruptions 
(e.g., \citeA{torok.et11,schrijver.et13,sterling.et14}). These aspects of AR jets, in comparison to QS and
CH jets, have yet to be fully understood.

Regarding the magnetic source of AR jets: Although  \citeA{sterling.et16b} and \citeA{sterling.et17} found
all but one of their jets to occur at sites of unmistakable flux cancelation, in several cases these
cancelations occurred in flux emergence regions, where one pole of the emerging flux ran into pre-existing
opposite-polarity flux, with the jet occurring along the so-formed neutral line.  In the 20 AR jets
they examined, \citeA{mulay.et16} report that 10 occurred due to flux cancelation, and four occurred
at sites of emergence and then cancelation; thus 14 out of 20 were due to flux cancelation.  They 
further report however that six of the 20 occurred with flux emergence without cancelation.  We point
out that in ARs the magnetic field can be very dynamic, and it can be difficult to ascertain with 
certainty whether flux emergence is truly occurring without cancelation of between one of the poles
and nearby opposite-polarity field (or even cancelation at the neutral line at the center of the
emerging bipole itself).  Further studies of jets, both inside and outside of ARs, should be carried 
out to confirm the magnetic source of the jets.

Although we do not detail spectroscopic studies of jets here (e.g, \citeA{chifor.et08,tian.et12}; see \citeA{raouafi.et16} 
for more citations), there is hope that resulting information regarding twisting motions, mass flows, temperatures, 
and densities can help clarify observational properties of jets in all solar regions, and give further insight into their
driving mechanism.

\subsection{Jet Numerical Models: Recent Progress}
\label{subsec-jet models}

\citeA{raouafi.et16} reviewed in detail earlier jet numerical simulation studies.  Since that time,
\citeA{liu.et16} presented numerical simulations based on the emerging-flux model, extending the 
simulations of \citeA{fang.et14} to describe a nearly concurrent pair of jets (twin jets). \citeA{ni.et17}
also used the flux-emergence model to explore blobs in jets resulting from the plasmoid and Kelvin-Helmholtz
instabilities.  As discussed above, there is a question of how common it might be for flux emergence to
be the primary trigger of jets.  But even if the jet trigger is something different, some of the physics 
described by these models should still be relevant to aspects of jet phenomena.

Several studies have not explicitly relied on flux emergence for producing jets.  One such set of studies by
Pariat and colleagues (\citeA{pariat.et09,pariat.et10,pariat.et15}, mentioned earlier)
assumes an anemone field embedded inside of an ambient background field, where the system is energized via a rotational
motion of the anemone field inside of the surrounding ambient field.  Eventual reconnection between the flux 
systems results in energy release and jets along the ambient field extending vertically above the bipole.
\citeA{karpen.et17} and \citeA{roberts.et18} extended these studies to consider, among other effects, the
manifestation of the resulting jets in the far corona.  \citeA{szente.et17} use a variation of the same model,
whereby the twisting anemone is located beneath the photosphere, and report close matches with physical 
structure, dynamics, and emission of observed EUV and X-ray jets.

So far only one set of studies has explicitly attempted to simulate the minifilament-eruption model
for jets.  \citeA{wyper.et17} use an initial simulated  magnetic setup like that of Figure~6a, and 
they are able to reproduce the process pictured in the subsequent panels of 
Figure~6, therefore providing numerical support for the concept, including the jet spire and the 
representation of the JBP as a miniature flare beneath the erupting minifilament flux rope.  \citeA{wyper.et18a} 
and \citeA{wyper.et18b} extend the simulations to differing environmental circumstances.  
They refer to their simulations as a ``breakout model for jets,'' because the reconnection at the magnetic null
point, in between the bipole field and the ambient overlying field (and where the external 
reconnection occurs at the upper red X in Fig.~6b), is identical to the breakout model setup developed 
for larger-scale eruptions \cite{antiochos98,antiochos.et99}.  

To date, no model has self-consistently produced jets based on minifilament eruptions triggered by 
magnetic flux cancelation.  As per the discussion preceding, several recent observations suggest
that this is how at least a large fraction of jets operate.  Some recent simulations give hope that
such simulations might be possible in the near future \cite{galsgaard.et19,syntelis.et19}.

AR jets (and probably, to some extent, non-AR jets as well) can be the source for accelerated particles. 
See \citeA{raouafi.et16} a discussions.  Examples of more recent work include \citeA{nitta.et15} and
\citeA{glesener.et18}.

\section{Jet-like Phenomena on Differing Size Scales}
\label{sec-extensions}

\subsection{Coronal Jets and Large-Scale Eruptions}
\label{subsec-jets and large eruptions}

If coronal jets are truly small-scale versions of larger-scale eruptions, then we would expect
the processes that trigger jets perhaps also to trigger large-scale eruptions. (In this case, we are
referring to large-scale flux-rope eruptions that produce a flare in tandem with a CME, or 
a confined flare eruption without a CME\@.)  Because of the 
substantial evidence that converging and canceling fields trigger jet eruptions, we can question 
whether they also trigger the large-scale eruptions.  A difference however between jets and the 
larger eruptions is that the magnetic-field structures that produce jets (specifically, minifilaments) 
form only hours to days before onset of typical jets, based on the work of \citeA{panesar.et17}.  In contrast,
the build up to large-scale eruptions can persist for several weeks before the eruptions
occur, and therefore longer than a two-week disk-passage period.  Therefore it is much easier to 
follow the complete life span of the magnetic structures of small-scale jets compared to larger-scale 
erupting regions.

In order to determine with more confidence whether large-scale eruptions mimic coronal jets (i.e., the
formation and dynamics of minifilaments that erupt to form jets), \citeA{sterling.et18} looked at two
ARs that were small enough so that they could be followed from the time of their birth until the time
of of the CME-producing eruptions. They found that, in both cases, the regions emerged as bipoles,
spread apart, and then the two polarities approached each other again, with the entire evolution
lasting roughly five days in each case.  As the two poles converged, they clearly underwent cancelation
at the neutral line at the center of the respective bipoles.  This apparently built flux ropes (or sheared-field 
filament channels), along
which filaments formed.  These filaments erupted, producing CMEs.  In one of the regions, an additional
filament formed on a neutral line between one of the bipoles spreading during emergence and an exterior
ambient opposite-polarity field; that filament also erupted.  Other then this, the regions were 
magnetically isolated, so that they had little other interaction with surrounding field.   This behavior 
is similar to that of jet magnetic elements, in which both the minifilament
is formed by canceling fields and the eruption is later triggered by canceling fields (e.g., 
\citeA{panesar.et16a,panesar.et17}).

Some flux-rope CMEs occur from eruptions that do not have an obvious erupting cool-material filament; 
in that case, the
field still erupts and launches a flux rope outward, but for whatever reason a filament had not formed on the
field prior to its eruption.  Thus we might expect that some jets also occur without a discernible cool 
minifilament (e.g., \citeA{sterling.et17}).  Studies of the percentages of jets that do or do not carry 
such erupting cool filament material are currently underway.

\subsubsection{Coronal Jets and CMEs}
\label{subsubsec-jets and CMEs}

Thus the erupting minifilament in jets resembles the eruptions that cause CMEs in larger events.  In
jets, the erupting minifilament can also produce ejections that extend into the heliosphere.  
We discuss two different mechanisms for forming CMEs from jets.

One mechanism produces the narrow features that are called ``white-light jets"
and observed in space-borne coronagraphs.  These are also called ``narrow CMEs,''
as they typically have angular widths (measured from Sun center) of less than 10$^\circ$ 
or 15$^\circ$, with different studies using different criteria; this compares with an
average width of $\sim$44$^\circ$ for all CMEs \cite{gopal.et09}. There are several other examples of 
such narrow CMEs from jets (e.g., \citeA{wang.et98,wang.et02,gilbert.et01,dobrzycka.et03,yashiro.et03,
bemporad.et05,nistico.et09,hong.et11,shen.et12,moore.et15,sterling.et16b}).
\citeA{wang.et98} showed that the source of these white-light jets on the Sun was coronal jets (that they observed
with EIT\@). \citeA{moore.et15} showed that the coronal jets that displayed the strongest 
untwisting motions tended to show up as narrow CMEs in \soho/LASCO coronagraph images.
Moreover, they showed evidence that the jet twisting motions propagated onto the white-light
jets, manifesting as swaying-like dynamics of those coronagraph structures.  

In terms of the minifilament-eruption model, this white-light-jet twisting can be understood 
as discussed in \S\ref{subsec-magnetic causes}, that is,
by imagining that the erupting minifilament contains twist, and that twist is then transferred 
onto the ambient open field via the external reconnection of Figure~6b, 6c.  Furthermore,
this provides a ready explanation for the difference between a white-light jet and a 
conventional CME: In the white-light jet, the erupting flux rope (that is, the magnetic-flux-rope
shell surrounding the erupting cool-material minifilament) contains a relatively small amount
of total flux, and all of this flux is transferred to the ambient open field; that is, the
minifilament-flux rope is totally consumed, or ``eaten up,'' by the ambient field.  In contrast,
for a typical CME, the flux contained in the erupting flux rope is large enough (compared to the
surrounding field) so that a substantial portion of the flux rope can escape the corona 
intact, appearing as the bubble portion of the CME core.

The second mechanism we describe for producing CMEs from jets results in what have been called ``streamer puff''
CMEs \cite{bemporad.et05,alzate.et16,panesar.et16b}.  In this case, a jet occurs at one side of  
the base of a coronal streamer, and a loop in the ``helmet" of the streamer is blown out.  
\citeA{bemporad.et05} provided an explanation for these features, but this explanation was updated
by \citeA{panesar.et16b} to mesh with the minifilament-eruption picture of jets.  This explanation says that,
instead of undergoing external reconnection with an ambient open field as in the case of the narrow-CMEs, 
an erupting twisted minifilament field instead undergoes reconnection with the 
streamer-helmet loop field, rendering the streamer loop unstable so that it erupts outward as
the streamer-puff CME\@.

\subsection{Smaller-scale Jets: Jetlets and Spicules(?)}
\label{subsec-small jets}


\citeA{sterling.et16a} suggested that there might be a power law relationship for eruptions of filament-like
structures, with CME-producing large-scale filament eruptions on the large end, extending down through 
coronal jets, and possibly continuing down to spicules on the small end.  We emphasize that filament-like features have not
been observed in spicules, and therefore extending this concept to spicules rests on the speculative assumption that spicules
result from eruptions of so-far-unseen {\it microfilaments}.  With that caveat, \citeA{sterling.et16a} plotted estimates for the
number of spicules, coronal jets, and CME-producing eruptions that are on the Sun at a given at a given time,
against the size (observed or estimated) of the erupting filament-like feature.  This produces the plot of
Figure~8, which we show with a linear fit line overlaid.  For the estimate of the number of spicules, 
they used published numbers of spicule occurrences from \citeA{athay59} and \citeA{lynch.et73}. A recent
study by \citeA{judge.et10} however suggests that these values may be severely underestimated, and estimate 
that there are $\sim$2$\times 10^7$ spicules on the Sun at any given time.  Therefore the uncertainty
in the spicule counts is very high, and the plotted point in Figure~8 should be regarded as a minimum value.
Therefore it is possible that only some percentage of spicules are jet-like in nature.  Those spicules would
be analogous to coronal jets, and the formation of those spicules would be just as in the pictured
presented in Figure~6, but scaled down to occur only in the photosphere and chromosphere.  Moreover, these
spicules would be expected to form at very weak neutral lines, and likely where magnetic cancelation is occurring.  
But how big or small is the percentage of spicules formed this way is currently not known, and possibly can 
only be addressed with better observations, such as with the upcoming DKIST telescope.

\begin{figure}[ht!]
\hspace*{0.2cm}\includegraphics[angle=0,scale=0.70]{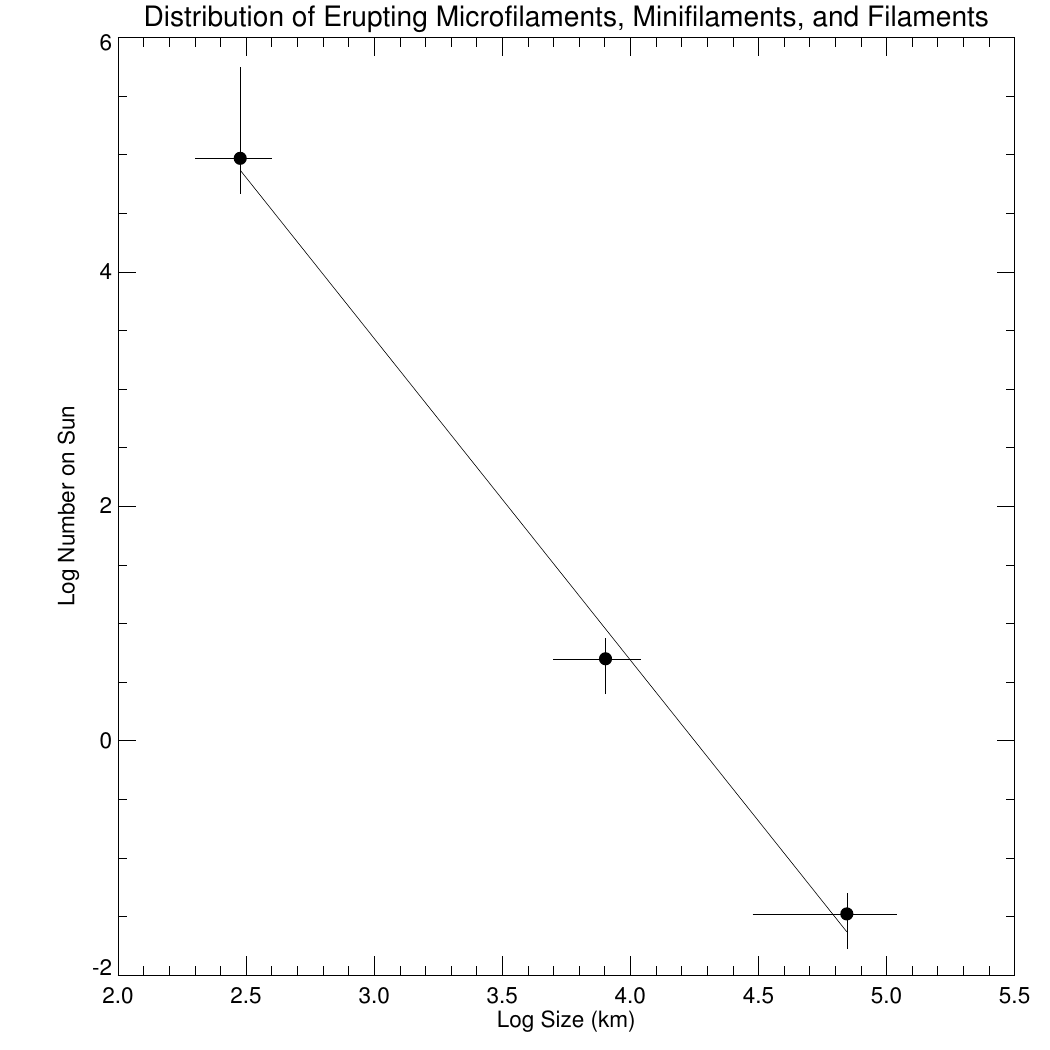}\vspace{0.0cm}
\caption{Plot of estimated observed number of erupting filament-like features on the Sun at any given time (vertical axis) as a function
of the size of the feature.  The rightmost point represents CME-producing filament eruptions; the middle point
represents observed coronal jets that result from minifilament eruptions; and the leftmost point represents spicules, 
under the still-speculative assumption that they result from eruptions of postulated {\it microfilaments}, of size 
similar to spicule observed widths, and with a number-on-Sun value matching estimates of 
\citeA{athay59} (and with the top of that 
point's vertical bar matching the higher estimate due to \citeA{lynch.et73}).  As 
mentioned in the text, the number of spicules may be underestimated, perhaps severely so.  The line is a least-squares fit
to the three points. This figure is 
from \citeA{sterling.et16a}; see that paper for further details. \label{fig8}}
\end{figure}

There is however evidence for jet-like features that, while not as small as spicules, are smaller than typical
jets, and which can be resolved with current instruments.  These features were called {\it jetlets} by
\citeA{raouafi.et14}.  They identified them as transient jet-like features occurring at the sites of
quasi-random magnetic cancelations at the bases of plumes in coronal holes.  They proposed that the jetlets,
along with plume transient bright points (PTBPs), help sustain the plumes over their lifetime of several days.

\citeA{panesar.et18b} confirmed the presence of jetlets in plumes, but also found that they are prevalent in 
non-plume chromospheric network regions.  Hence, they refer to the features with the more general term {\it network
jetlets}; these are similar to or the same as the network jets of \citeA{tian.et14} 
(also see, e.g., \citeA{narang.et16,kayshap.et18}).  Using AIA/EUV and \iris/UV images, for 10 measured jetlets, 
\citeA{panesar.et18b} found them to have 
lengths $\sim$30,000~km, spire widths 3000~km, and durations of three~min. They also found that nine of them 
occurred at sites of magnetic cancelations; while the origin of the tenth was uncertain, they speculate that it too 
was due to (unresolved) flux cancelation.

\section{Coronal Jets and Other Phenomena}
\label{sec-other phenomena}

We close by addressing briefly other important topics on jets, but which are not addressed in detail here.

\subsection{Jets and Plumes}
\label{subsec-plumes}

As noted in \S\ref{subsec-small jets}, \citeA{raouafi.et14} suggested that jetlets and PTBPs might sustain plumes over
their several-day lifetime.  Even earlier however, \citeA{raouafi.et08} and \citeA{raouafi.et10} used \stereo/SECCHI and
\hinode/XRT data to present evidence that the formation of plumes is preceded by coronal jets occurring at the same
locations. \citeA{pucci.et14} also found that a long-lived plume they studied was preceded by a coronal jet. While 
the idea that jets are critical for plume formation requires further verification \cite{raouafi.et14}, it is an important
consideration in understanding the source of plume material, and also the magnetic environment at the base of plumes.
See \citeA{poletto15} for further discussions.

\subsection{Jets and Macrospicules}
\label{subsec-macrospicules}

\citeA{pike.et97} claimed that X-ray jets and EUV macrospicules were manifestations of the same phenomenon 
(cf.\ \citeA{kiss.et17}).  Subsequent observations support this view, at least for some X-ray jets 
(\citeA{sterling.et10a,curdt.et12,moore.et10,moore.et13}).  Combined with the minifilament-eruption view
of X-ray jets, this supports that macrospicules are the result of miniature filament eruptions, as proposed
by \citeA{moore.et77} and \citeA{labonte79}.

\subsection{Jets and Coronal Heating}
\label{subsec-coronal heating}

Clearly coronal jets supply energy to the corona, but so far it is unclear whether jets (on various size scales) can 
supply enough energy to explain general coronal heating.  Individual jets are estimated to contribute something like
$10^{26}$---$10^{27}$~erg for coronal hole jets, and $10^{28}$---$10^{29}$~erg for AR jets \cite{shimojo.et00,pucci.et13,
sterling.et17,sterling.et18}. Considering coronal hole jets in aggregate, \citeA{paraschiv.et15} estimated the amount of heating due
to jets to be $\sim$2$\times 10^3$~erg~cm$^2$~s$^{-1}$ (comparable to values obtained by \citeA{yu.et14}), 
which is far below, e.g., the rate of 6$\times 10^5$~erg~cm$^2$~s$^{-1}$ required for coronal holes \cite{withbroe.et77}.  
\citeA{poletto.et14} and \citeA{szente.et17} also found energy rates from jets less than that required to sustain coronal heating, based
on the estimated contribution of coronal jets alone.  \citeA{moore.et11} argue however that the heating rate could be 
sufficient for coronal heating, provided that the jet process continues down to events the size scale of spicules, 
and if spicules result from erupting magnetic bipoles as described in that \citeA{moore.et11} paper.

\section{The Future} 
\label{sec-future}

The study of solar jet-like structures is clearly rich and fascinating, with these features forming fundamental components
of the solar atmosphere.  As demonstrated above, whether spicules, coronal jets, or other features such as jetlets, 
more work is required to confirm what mechanism (or which mechanisms) is/are driving them.  New instrumentation in the
near future, such as DKIST and upcoming high-resolution space missions, will play a crucial role in unraveling these mysteries.

\acknowledgments

This work was supported by funding from the Heliophysics Division of NASA's
Science Mission Directorate  through the Heliophysics Guest Investigators (HGI) Program, and 
the NASA/MSFC \hinode\ Project.


\bibliography{ms}


\clearpage


\end{document}